\documentclass[titlepage,onecolumn,11pt,nofootinbib,prd,floatfix,preprintnumbers,amsmath,amssymb,groupedaddress,superscriptaddress]{revtex4}
\usepackage{epsfig}  
\usepackage{array}  

\begin{document}   
\title{Anomaly Holography} 
\date{14th February, 2008 }
\author{Ben Gripaios}  
\email{b.gripaios1@physics.ox.ac.uk}  
\affiliation{Rudolf Peierls Centre for Theoretical Physics, University of Oxford,
1 Keble Rd., Oxford OX1 3NP, UK}
\affiliation{Merton College, Oxford OX1 4JD, UK}
\author{Stephen M. West}  
\email{s.west1@physics.ox.ac.uk}  
\affiliation{Rudolf Peierls Centre for Theoretical Physics, University of Oxford,
1 Keble Rd., Oxford OX1 3NP, UK}
\begin{abstract} 
We consider, in the effective field theory context, anomalies of gauge field theories on a slice of a five-dimensional, Anti-de Sitter
 geometry and their four-dimensional, holographic duals. A consistent effective field theory description 
 can always be found, notwithstanding the presence of the anomalies and
 without modifying the degrees of freedom of the theory. If anomalies do not vanish,
the $d=4$ theory contains additional pseudoscalar states, which
 are either present in the low-energy theory as physical, light states, or are eaten by (would-be massless) gauge bosons. 
We show that the pseudoscalars ensure that global anomalies of the four-dimensional dual satisfy the 't Hooft matching condition and comment on the 
relevance for warped models of electroweak symmetry breaking.
\end{abstract}   
\preprint{OUTP-0705P}
\maketitle 
\section{\label{intro}Introduction}
Maldacena's AdS/CFT correspondence \cite{Maldacena:1997re}, which has now been around for a decade or so, 
has provided us with
remarkable insights into strongly-coupled gauge theories, and has passed numerous consistency checks. 
One such check, pointed out by E. Witten \cite{Witten:1998qj}, is that Chern-Simons
terms in the action on the AdS side, which are not gauge-invariant on the projective boundary of AdS, 
correctly reproduce the anomaly structure of the global $R$-symmetry currents on the CFT side.

It is believed that the correspondence is rather general and can, in particular, be extended to cover the situation
where the AdS space is truncated by one or more branes \cite{Arkani-Hamed:2000ds,Rattazzi:2000hs,Perez-Victoria:2001pa}.
 On the CFT side, this is interpreted as a breaking of the 
conformal symmetry. If the theory on the AdS side contains fermions, the presence of the 
branes leads to additional, anomalous
contributions to gauge transformations in the $d=5$ theory. These contributions have already been discussed for a flat extra-dimensional
geometry, for example, in, \cite{Callan:1984sa,Arkani-Hamed:2001is,Scrucca:2001eb,Barbieri:2002ic,vonGersdorff:2003dt,Scrucca:2004jn,Boyarsky:2005eq} and for a warped 
geometry in \cite{Hirayama:2003kk}; a clarification of these contributions
and the r\^{o}les they play in the effective field theories
on both sides of the AdS/CFT correspondence will be the goal of this Paper. We will also discuss the relevance
for warped models of electroweak symmetry breaking.

We consider the correspondence in its crudest form, in which 
a gauge theory in a slice of an $AdS_5$ space (the so-called RS1 geometry \cite{Randall:1999vf}),
weakly-coupled to itself and to gravity, is dual\footnote{Up to energy scales
of order of the AdS curvature scale, where the volume of the internal manifold, $M$, of the full $AdS \times M$ 
geometry, is resolved.} to some gauge theory in four dimensions with large rank and strong coupling.
This four-dimensional theory has a conformal invariance that is broken explicitly 
by an ultra-violet (UV) cut-off and is non-linearly realized in the infra-red (IR).

We find that a consistent effective field theory (EFT) description
can always be found, notwithstanding the presence of anomalies.
In seeking this description, we take great care not to modify the existing degree-of-freedom content of the theory.
The resulting low-energy spectrum in $d=4$ 
is not that which might be obtained by a na\"{\i}ve dimensional reduction of the $d=5$ theory. Indeed, we find that
a consistent description generically contains extra, scalar states\footnote{Actually, these are pseudoscalars but we do not make the distinction in what follows.}. 
These are either present in the low-energy effective
theory as light, physical states, or are eaten by gauge bosons, producing a theory with fewer massless gauge bosons in the low-energy spectrum.
These extra scalar states play an essential r\^{o}le in the AdS/CFT correspondence: 
they lead to Wess-Zumino-Witten (WZW) terms
in the low-energy effective action that guarantee that global anomalies in the $d=4$ dual theory obey the 't Hooft matching condition
\cite{tHooft:1980xb} at all energy scales.

Models of this type, in which a local symmetry group $G$ in the bulk of $AdS_5$ is broken to subgroups $H_0$ and 
$H_1$
on the branes (located at positions $z_0$ and $z_1$, respectively, in the fifth-dimensional co-ordinate)
have been extensively invoked as models of 
electroweak symmetry breaking (EWSB), both with \cite{Contino:2003ve,Contino:2006qr} and without 
\cite{Csaki:2003zu} Higgs scalars. They provide natural models in which the hierarchy between the Planck and electroweak scales
is explained either by the warp-factor of the AdS geometry on the $d=5$ side, or equivalently by the `slow' running of the coupling
constants on the $d=4$ side. 
The effective theory at, or below, the weak scale is that of an $H_0$-gauged $G/H_1$ non-linear sigma model,
with the massive gauge bosons corresponding to generators in $H_0$, but not in $H_1$, integrated out.
We show that specifying a plausible fermion content for one such model, the so-called MCHM$_5$, renders it manifestly
anomaly-free. Doing the same for other models results in non-vanishing fermion anomalies, which will need to be cancelled by adding Chern-Simons
terms or, more generally, altering the fermion content; such changes, of course, affect
 the phenomenology, which we hope to explore in future work \cite{gmw}.

Although our findings apply to the general case of arbitrary bulk gauge group $G$ broken to subgroups
$H_{0,1}$ on the branes, most of the pertinent features are already extant in the simplest case where $G = U(1)$, 
and we shall use this as our primary example in the sequel.

The outline is as follows. In the next Section we review the contributions to gauge-variance in $d=5$, coming from Chern-Simons terms and fermionic anomalies, and their
relevance in the EFT context. In Section III, we discuss the $d=4$ holographic interpretation. In Section IV we discus the implications for models of EWSB.

The work presented here overlaps with a recent preprint \cite{Panico:2007qd}, which, in particular, describes
the holographic connection between the $d=5$ Chern-Simons term and WZW terms in $d=4$.

\section{\label{d=5}Gauge-variance in d=5}

Consider a gauge theory on a $d=5$ spacetime that is topologically of the form $\mathbb{R}^4 \times I$, where $I$ is
a closed interval of the real line. We use co-ordinates $x^{\mu}$ on $\mathbb{R}^4$ and $z$ on the interval $I=[z_0,z_1]$.
We refer to the disjoint boundaries of the interval as the $z_0$, or UV, brane and the $z_1$, or IR, brane.
Equivalently \cite{Csaki:2005vy}, one can think of a space obtained as some orbifold of the space $\mathbb{R}^4 \times S^1$. We shall always
use the interval formulation however, since it is more convenient for dealing with the general boundary conditions (BCs)
that we shall employ. 

Imposing the requirement of gauge-invariance in the bulk does not necessarily imply gauge-invariance on the branes, either at the 
classical or quantum level. Indeed, consider a Chern-Simons term in the action for the U(1) theory in the bulk of the form\footnote{
We use latin majuscules for $d=5$ indices, e.g. $M \in \{0,1,2,3,5\}$, and greek minuscules for $d=4$, e.g. $\mu \in \{0,1,2,3\}$.
Our conventions for Dirac matrices, {\em \&c.} are those of \cite{Gripaios:2006dc}.}
\begin{gather}\label{CS}
S_{\mathrm{CS}} = c \int d^4x dz \;  \epsilon^{MNPQR}A_M F_{NP}F_{QR}.
\end{gather}
Under a $d=5$ gauge transformation of the form
\begin{gather} \label{gt}
A_M \rightarrow A_M + \partial_M \Lambda (x,z),
\end{gather}
we find
\begin{align}
\delta S &= c \int d^4xdz \;  \Lambda \epsilon^{5\mu\nu\rho\sigma} F_{\mu\nu}F_{\rho\sigma}
\Big[-\delta (z-z_0)+ \delta (z-z_1)\Big], \nonumber \\
&= 2c \int d^4xdz \;  \Lambda F_{\mu\nu}\tilde{F}^{\mu \nu}\Big[-\delta (z-z_0)+ \delta (z-z_1)\Big],
\end{align}
where $\tilde{F}^{\mu \nu } = \frac{1}{2}\epsilon^{\mu\nu\rho\sigma} F_{\rho\sigma}$.
So the action is not gauge-invariant on the branes even at the classical level, in the presence of the Chern-Simons term.

This is not necessarily the only source of gauge-variance however. Charged fermions, either propagating in the bulk or
localized on the branes, can also lead to gauge-variance at the quantum level, via anomalies.

Consider first a bulk fermion of charge $Q$ with respect to the $U(1)$ gauge symmetry. The smallest irreducible representation
of the Dirac algebra in $d=5$ has dimension four and is carried by a Dirac spinor $\Psi$. In a non-compact theory,
this precludes the existence of a perturbative gauge anomaly, but this is not true in the presence of the branes, 
where different boundary conditions for the two Weyl spinors making up $\Psi$ make the theory intrinsically chiral,
 as is clear from the existence of chiral zero modes in the $d=4$ spectrum.

To see the resulting anomaly, let us take $(x^{\mu},z)$ as Poincar\'{e} co-ordinates on a slice of $AdS_5$
with curvature scale $k$, such that the metric (signature mostly-plus) is given by
\begin{gather}\label{adsmetric}
ds^2 = \frac{1}{(kz)^2}(\eta_{\mu \nu} dx^{\mu} dx^{\nu} + dz^2).
\end{gather}
The action for $\Psi$, including coupling to the gauge field, is
\begin{gather}
S = - \int d^4 x dz\; \sqrt{-g} \frac{i}{2}\Big[ \overline{\Psi}\Gamma^M D_M \Psi -  M \overline{\Psi}\Psi \Big] + \mathrm{H.\ c.},
\end{gather}
where $M$ is the bulk Dirac mass, $\Gamma^M = e^M_A \gamma^A$ are curved-space gamma matrices and $D_M = \partial_M + \omega_M -i Q A_M$ includes the spin 
connection. In terms of the re-scaled Weyl spinors, $\psi_{\alpha}$ and $ \overline{\chi}^{\dot{\alpha}}$, defined such that 
\begin{gather}
\Psi = (kz)^2 \begin{pmatrix} \psi_{\alpha} \\ \overline{\chi}^{\dot{\alpha}}  \end{pmatrix},
\end{gather} 
the action is
\begin{multline} \label{actpsi}
S  = - \int d^4 x dz \; \Big[ 
-i\chi \sigma^{\mu} (\partial_{\mu} -iQA_{\mu}) \overline{\chi}
-i\overline{\psi}\overline{\sigma}^{\mu} (\partial_{\mu} -iQA_{\mu}) \psi \\
+\frac{1}{2}\big( \chi (\partial_5 -iQA_5) \psi + (\partial_5 +iQA_5) \overline{\psi} \overline{\chi} - 
\overline{\psi}(\partial_5 -iQA_5) \overline{\chi} \\
-(\partial_5 +iQA_5)\chi \psi \big) +\frac{M}{kz}(\chi \psi + \overline{\psi} \overline{\chi}) \Big].
\end{multline}
Requiring that the variation of the action vanishes in the bulk gives the bulk equations of motion,
\begin{align} \label{bulkpsi}
0&= -i\overline{\sigma}^{\mu} (\partial_{\mu} -iQA_{\mu}) \psi - (\partial_5 -iQA_5) \overline{\chi} +\frac{M}{kz}\overline{\chi},\nonumber \\
0&=-i\sigma^{\mu} (\partial_{\mu} -iQA_{\mu}) \overline{\chi} - (\partial_z -iQA_z) \psi +\frac{M}{kz}\psi.
\end{align}
Requiring that the variation of the action vanishes on the branes gives the condition
\begin{gather}
0 =  \frac{1}{2}\int d^4 x \; \Big[
\delta\chi \psi - \delta\overline{\psi} \overline{\chi} +\delta\overline{\chi}\overline{\psi}-\delta \psi \chi
\Big]_{z_0}^{z_1},
\end{gather}
which is satisfied by BCs of the form \cite{Csaki:2003sh}
\begin{gather}
\chi_{\alpha} =N_{\alpha \dot{\beta}}{\overline{\psi}}^{\dot{\beta}},
\end{gather}
at $z=z_0,z_1$, where
$N_{\alpha}^{\dot{\beta}} = \pm {N^{\dagger}}_{\alpha}^{\dot{\beta}}$ ($N_{\alpha}^{\dot{\beta}}$ may include boundary derivatives, and the $\pm$ signs account for partial 
integrations of these.).

Let us focus on just a subset of the possible BCs, where either $\psi$ or $\chi$ is set to zero on each brane.
We label the various possibilities by the ordered pair $(\alpha_0,\alpha_1)$, where the first entry refers to the $z_0$ brane
and the second entry to the $z_1$ brane. The entries $\alpha_0$ and $\alpha_1$ take values in $\{+,-\}$; a 
plus indicates that $\chi$ vanishes on the relevant brane, whereas a minus indicates that $\psi$ vanishes. The reason
for this notation will become evident when we consider the anomalies arising from the bulk
fermion with the specified boundary conditions.

Before we do that, let us point out that \cite{Csaki:2005vy} the boundary conditions (+,+) give rise to a left-handed massless Weyl fermion
in $d=4$, the $(-,-)$ conditions give rise to a right-handed Weyl fermion, and the other possible BCs do not give rise to massless modes.

Returning to the anomalies, let us consider the effective action, $\Gamma [A_M]$, in $d=5$ obtained by integrating out the bulk fermion in
a $U(1)$ gauge field background, {\em viz.}
\begin{gather}
\exp i \Gamma [A_M] = \int D \overline{\Psi}D\Psi \; \exp i S[\Psi,A_M].
\end{gather}
As a result of the anomaly, $\Gamma [A_M]$ is not invariant under a background gauge transformation of the form (\ref{gt}).
The variation for the BCs $(\alpha_0,\alpha_1)$ is easily determined by comparison with results previously obtained for $S^1/\mathbb{Z}_2$
 \cite{Arkani-Hamed:2001is} and $S^1/\mathbb{Z}_2 \times {\mathbb{Z}}'_2$
orbifolds \cite{Scrucca:2001eb}, and is given by
\begin{gather} \label{an}
\delta \Gamma [A_M] = \int d^4 x d z \; \Lambda(x,z) \; \mathcal{A}(x,z),
\end{gather}
where
\begin{gather}
\mathcal{A}(x,z) = \frac{Q^3}{96 \pi^2} F_{\mu \nu} \tilde{F}^{\mu \nu}\Big[\alpha_0\delta (z-z_0)+\alpha_1\delta (z-z_1)\Big].
\label{bulkanom}
\end{gather}
Let us pause to examine this result. Firstly, we see that the anomalies are localized on the branes.
This should come as no surprise, since it is only via the boundary conditions that any notion of chirality is introduced.
The anomaly is a topological artefact, and for
 the same reason, the result is independent of the metric and the bulk mass $M$ of the fermion.
 Secondly, the numerical factor deserves comment. There is an extra factor of two in the denominator
 relative to the usual consistent anomaly of a $d=4$ Weyl fermion \cite{Hill:2006ei}. This extra factor of two indicates that the anomaly
 is split between the two branes: if we integrate $\mathcal{A}(x, z)$ over the $z$ co-ordinate, we find a $d=4$
 anomaly whose value is the consistent anomaly of a left-handed Weyl fermion, multiplied by a 
 factor $\frac{\alpha_0+\alpha_1}{2}$, which takes values $+1$, $-1$, or $0$ for
 the BCs $(+,+)$, $(-,-)$, or $(\pm,\mp)$ respectively. We see that this $d=4$ anomaly is in one-to-one correspondence
 with the anomaly produced by the $d=4$ massless fermion modes.
 
 Anomalies can also occur due to brane-localized fermions. Unsurprisingly, they result in a gauge-variation of the effective action
 (obtained in the usual way by integrating over the brane-localized fermions) that is localized on the relevant brane and whose value there is
 given by the usual $d=4$ anomaly. For example, a left-handed Weyl fermion of charge $Q$ on the $z_0$ brane results in a variation
\begin{gather}
\mathcal{A}(x,z) = \frac{Q^3}{48 \pi^2} F_{\mu \nu} \tilde{F}^{\mu \nu}\delta (z-z_0).
\end{gather}

So there are three possible contributions to gauge-variance in theories of this type, {\em viz.} from
Chern-Simons terms for bulk gauge fields, and from bulk and brane-localized fermions. 
All contributions are localized on the branes and, in a convenient abuse of notation, 
we shall refer to them collectively as brane-localized anomalies (though the classical variation of
the Chern-Simons term is not a quantum anomaly).

None of these observations is new, and indeed the implications
of the brane-localized anomalies for physics have already been discussed at length in the
literature \cite{Callan:1984sa,Arkani-Hamed:2001is,Scrucca:2001eb,Barbieri:2002ic,vonGersdorff:2003dt,Scrucca:2004jn,Boyarsky:2005eq}.
In the original work \cite{Arkani-Hamed:2001is} (formulated on a flat $S^1/\mathbb{Z}_2$ orbifold),
it was shown that the usual $d=4$ anomaly cancellation condition, applied to the $d=4$ fermion zero modes,
was a sufficient condition for cancellations of the $d=5$ brane-localized anomalies. But this is not true
for a more general orbifold (as was pointed out for the orbifold $S^1/\mathbb{Z}_2\times {\mathbb{Z}}'_2$ in \cite{Scrucca:2001eb}).
Insufficiency is easily exhibited in the interval formulation by means of a counter-example: consider
a bulk fermion, charged under a bulk $U(1)$ gauge symmetry,
with $(+,-)$ BCs. Such BCs do not admit a chiral zero mode, so there is no $d=4$ anomaly. However, the $d=5$
brane-localized
anomalies are non-vanishing, albeit equal and opposite. 

In our counter-example (and in the example of \cite{Scrucca:2001eb}), the brane-localized anomalies can be cancelled,
without modifying the degrees of freedom of the theory,
by a Chern-Simons term with an appropriate coefficient \cite{Barbieri:2002ic}, rendering both the $d=4$ and $d=5$
theories anomaly-free. But it is easy in the interval formulation to construct a different counter-example, where anomaly cancellation cannot
be achieved by Chern-Simons terms: consider instead
bulk gauge group $G=SU(2)$, broken on both branes to the $U(1)$ subgroup generated by $T_3$. One possibility for the $d=5$ fermion content is to put a brane-localized 
left-handed Weyl fermions on each brane but with opposite $U(1)$ charge. There is no net anomaly in the spectrum of
fermion zero-modes, yet the equal and opposite anomalies on the branes cannot be cancelled by the non-Abelian Chern-Simons term, 
\begin{widetext}
\begin{gather}
S_{\mathrm{CS}} = c \int d^4x dz \;  \epsilon^{MNPQR}Tr(A_M\partial_NA_P\partial_Q A_R -\frac{3i}{2}A_MA_NA_P\partial_QA_R-\frac{3}{5}A_MA_NA_PA_QA_R),
\end{gather}
\end{widetext}
which  is proportional to $Tr(T^a \{T^b, T^c\})$ and therefore vanishes for $SU(2)$. Alternatively, we can take the fermion content to be a bulk fermion 
transforming as a doublet under the bulk $SU(2)$. The BCs for the $SU(2)$ doublet $\big(\begin{array}{cc}\Psi^1 & \Psi^2 \end{array}\big)^T$ need only respect the residual 
$U(1)$ symmetry on the branes; we take them to be
 $(+,-)$ for $\Psi^1$, which has charge $+1$ under the $U(1)$, and $(-,+)$ for $\Psi^2$, which has charge $-1$.
 Computing the anomaly as above, we find
\begin{gather}
  \mathcal{A} (x,z)  = 2 (\delta (z-z_0) - \delta (z-z_1)).
\end{gather}
Now, there are no massless fermion modes in $d=4$ and hence no $d=4$ anomaly. But, yet again, we have non-vanishing brane-localized anomalies
in $d=5$ that cannot be cancelled by a Chern-Simons term. 

That Chern-Simons terms do not suffice to cancel the brane-localized anomalies in $d=5$, given the the $d=4$ zero mode
anomalies vanish, was already observed in \cite{vonGersdorff:2003dt} for orbifold theories in arbitrary dimensions, with the Chern-Simons terms
generalized to four-form Green-Schwarz \cite{Green:1984sg} fields.

Finally, it was noted that $U(1)$ anomalies of the type occurring in our $SU(2)$ counter-example could be cancelled
by the addition of Green-Schwarz two-form bulk fields \cite{GrootNibbelink:2003gb}, or twisted Ramond-Ramond brane-localized fields \cite{Ibanez:1998qp}. But in adding such 
fields, it would appear, at least na\"{\i}vely, that the degree-of-freedom content of the theory is being changed.

Instead of trying to cancel the brane-localized anomalies in this way, we should like to follow a different tack,
motivated by the knowledge that we are dealing with theories in $d=5$, that are inherently non-renormalizable. 
They can, at best, be considered as EFTs,
valid up to some UV cut-off. In a renormalizable theory, the reasons for requiring anomaly cancellation are two-fold:
anomalies spoil both renormalizability and unitarity. In a non-renormalizable theory, as Preskill 
\cite{Preskill:1990fr} has pointed out,
the only relevant issue is: what is the cut-off scale, beyond which strong-coupling, unitarity violation, or other problems occur?
He has, moreover, given lucid arguments that show that one can {\em always} find a consistent EFT description of an anomalous gauge theory, valid up to some non-vanishing 
cut-off, provided one allows the anomalous local symmetries to be non-linearly realized.
Thus we should expect that a consistent description of an anomalous theory in $d=5$ can always be found, irrespective of whether
or not anomalies can be made to cancel, and without changing the degrees of freedom of the theory. 
Our principal aim will be to find this description.

What is more, there are two arguments that suggest that this approach will be instructive in the context of the AdS/CFT
correspondence. Firstly, the coefficient of the Chern-Simons term is a measure of the fermion content of the CFT \cite{Witten:1998qj}.
We should like to consider an arbitrary CFT, so we should also consider an arbitrary coefficient for the Chern-Simons term,
in which case the brane-localized anomalies will, in general, be non-vanishing.

Secondly, the AdS/CFT correspondence tells us that the $z$ co-ordinate in $AdS_5$ corresponds, roughly speaking,
to the energy scale in the $d=4$ dual. What is more, the branes at $z_0$ and $z_1$ somehow correspond to the UV and IR
of the $d=4$ dual, respectively. So the localization of the anomalies on the branes in $d=5$ should somehow encode information
about the anomalies in the UV and IR of the $d=4$ dual. What we shall find is rather satisfying: 
the consistency of the EFT description in $d=5$ ensures that the global anomalies of the $d=4$ dual satisfy the 't Hooft matching 
condition at all energy scales.

How, then, do we find a consistent EFT description of a theory in $d=5$ with brane-localized gauge anomalies,
without changing the degrees of freedom of the theory? 
The problem with anomalies is that they lead to a loss of gauge-invariance. In a gauge-invariant theory, on the other hand,
constructing a consistent EFT description is straightforward, because gauge invariance furnishes us with a set
of equivalent descriptions of the theory, any of which can be invoked as one's whim dictates.
Then, for example, the strong-coupling scale (at which calculability is lost) is easily determined by power counting
in a gauge in which the propagator is well-behaved, {\em e.g.} the 't Hooft-Feynman gauge. Then it is easy to
see that the strong-coupling scale is the true cut-off, because Lorentz invariance and unitarity are easily exhibited 
up to the strong-coupling scale, by working in a covariant gauge or unitary gauge, respectively.

But in an anomalous theory, the gauge-invariance, and its associated benefits, seem to be absent.
All is not lost, however, once we realize that gauge symmetry is not really a symmetry at all,
but rather (in the sense discussed above) a redundancy, a set of equivalent descriptions.
It is, furthermore, a redundancy that is easily resurrected, by adding dynamical scalar fields
transforming under the gauge group, and including terms involving the scalar fields that
cancel the original anomalous variations coming from the fermions. To show that the new description 
with additional scalar fields and non-anomalous gauge symmetry is equivalent to the old one, it suffices
(at least locally) to choose the gauge in which the scalars vanish. We are then left with a theory without
the scalars and without the gauge symmetry, {\em viz.} the original description. 
We stress that adding the scalars in this way does not change the degree-of-freedom content of the theory.

Following Preskill \cite{Preskill:1990fr}, let us see how this works in the case of a $U(1)$ gauge theory in $d=4$, coupled
to a left-handed Weyl fermion of charge $Q$. Under a background gauge transformation $A_{\mu} \rightarrow A_{\mu} 
+ \partial_{\mu} \lambda (x)$, the effective action obtained by integrating out the fermion varies, analogously to 
(\ref{an}) and (\ref{bulkanom}), as
\begin{gather}
\delta \Gamma [A_{\mu}] = \int d^4 x  \; \lambda 
\frac{Q^3}{96 \pi^2} F_{\mu \nu} \tilde{F}^{\mu \nu}.
\end{gather}
To resurrect the gauge invariance, introduce a scalar field $\theta$, transforming as $\theta \rightarrow \theta + \lambda$
under the $U(1)$, together with a non-renormalizable term in the Lagrangian of the form
\begin{gather}
\mathcal{L} \supset - \frac{Q^3}{96 \pi^2} \theta F_{\mu \nu} \tilde{F}^{\mu \nu}.
\label{restore}
\end{gather}
It is then apparent that gauge invariance of the effective action is restored, and further that the original
description is recovered by the gauge choice $\theta =0$. This is not the end of the story, however;
in the spirit of EFT, we should include all terms in the effective Lagrangian consistent with the symmetry,
since they will be generated by quantum effects, even if we do not include them {\em a priori}.
In particular, kinetic terms for $\theta$ will be induced, at leading order in the derivative expansion.
Thus, the full effective Lagrangian takes the form

\begin{gather}
\mathcal{L} = -\frac{1}{4g^2}F_{\mu \nu}F^{\mu \nu}
-i\overline{\psi}\overline{\sigma}^{\mu} (\partial_{\mu} -iQA_{\mu}) \psi 
+ \frac{f^2}{2}(\partial_{\mu}\theta - A_{\mu})^2
- \frac{Q^3}{96 \pi^2} \theta F_{\mu \nu} \tilde{F}^{\mu \nu}+ \dots,
\end{gather}
where the ellipsis denotes terms suppressed by powers of the scale $f$. Because of the gauge invariance,
the theory is manifestly Lorentz-invariant and unitary, up to the cut-off scale at which strong-coupling occurs,
which we estimate on the basis of na\"{\i}ve dimensional analysis to be around $4\pi f$. 
To go back to the original description, we set $\theta=0$; we see that we have an effective theory of a 
gauge boson of mass $gf$, coupled to a massless Weyl fermion, and valid up to a cut-off $4\pi f$.
In the gauge-invariant description, the massive gauge boson arises because the gauge symmetry is non-linearly realized
by the Goldstone boson $\theta$.

Having seen how the resurrection of gauge invariance is used to construct a consistent EFT in $d=4$, let us apply the same
idea to construct a consistent EFT of an anomalous gauge theory on an interval in $d=5$.
Consider again a $U(1)$ gauge symmetry in the bulk, unbroken on the branes at the classical quadratic level, but
with gauge-variant contributions from arbitrary
brane-localized anomalies of the form
\begin{gather}
\label{first}
\delta \Gamma [A_M] = \int d^4 x d z \; \Lambda \;\frac{Q^3}{192 \pi^2} \; F_{\mu \nu} \tilde{F}^{\mu \nu}
\Big[\alpha_0\delta (z-z_0)+\alpha_1\delta (z-z_1)\Big].
\end{gather}
The theory is gauge-invariant in the bulk, but not, in general, on the branes.
To make a manifestly-consistent EFT, we resurrect the gauge symmetry on the branes by adding brane-localized scalars,
$\theta_0$ and $\theta_1$, transforming as
\begin{align}
\theta_0 &\rightarrow \theta_0 + \Lambda(x,z_0), \nonumber \\
\theta_1 &\rightarrow \theta_1 + \Lambda(x,z_1),
\end{align}
together with brane-localized interaction terms of the form
\begin{gather}
S \supset - \int d^4 x d z \;  \frac{Q^3}{192 \pi^2}\; F_{\mu \nu} \tilde{F}^{\mu \nu}
\Big[\alpha_0 \theta_0\delta (z-z_0)+\alpha_1 \theta_1 \delta (z-z_1)\Big].
\end{gather}
Yet again, the spirit of EFT demands that we write down all terms consistent with the bulk $U(1)$ and other symmetries.
Up to quadratic order in the derivative and field expansion (and disregarding the fermions), we have
\begin{gather}
S =  S_{\mathrm{bulk}}+S_0 +S_1,
\end{gather}
where
\begin{align}
\label{second}
S_{\mathrm{bulk}} &= \int d^4 x d z \; \frac{1}{(kz)^5}\Big[-\frac{1}{4g^2}F_{MN}F^{MN} + \dots\Big], \nonumber \\
S_0 &= \int d^4 x d z \;\delta(z-z_0)\Big[-\frac{1}{4g_0^2}F_{\mu \nu}F^{\mu \nu} + \frac{f_0^2}{2}(\partial_{\mu} \theta_0-A_{\mu})^2 + \dots \Big],\nonumber \\
S_1 &= \int d^4 x d z \;\delta(z-z_1)\Big[-\frac{1}{4g_1^2}F_{\mu \nu}F^{\mu \nu} + \frac{f_1^2}{2}(\partial_{\mu} \theta_1-A_{\mu})^2 + \dots \Big].
\end{align}
Now the scales $f_0$ and $f_1$ are given by the natural scales on the respective branes, {\em viz.} $1/z_0$ and $1/z_1$,
 and $g_0$ and $g_1$ are the couplings of brane-localized gauge kinetic terms.

We note that, at the quadratic level (which determines the spectrum), the only effect of non-vanishing brane-localized
anomalies is to force the inclusion of brane-localized scalars in order to maintain a fully gauge-invariant description.
Now, the effect of adding brane-localized scalars is well-known \cite{Csaki:2005vy}; it changes the spectrum of zero modes in
$d=4$, when we do the Kaluza-Klein expansion. To see this, we first need to fix the gauge. Let us define the $\xi$-gauge
by adding to the bulk action the gauge-fixing term
\begin{gather}\label{gfterm}
S^{\mathrm{GF}}_{\mathrm{bulk}}= -\int d^4 x d z \; \frac{1}{2\xi kzg^2}\Big[\partial_{\mu}A^{\mu}+\xi z \partial_5\left(\frac{A_5}{z}\right)\Big]^2.
\end{gather}
The gauge is not fixed completely. There are residual gauge transformations $\Lambda (x,z)$, such that
\begin{gather}
\partial_{\mu}\partial^{\mu} \Lambda + \xi z \partial_5 \left(\frac{\partial^5 \Lambda}{z} \right)=0,
\end{gather}
under which (\ref{gfterm}) is invariant.
This is a second-order differential equation in $z$, whose solution contains two arbitrary functions of $x$.
To fix these two, residual, $d=4$, gauge symmetries, 
we also define the $\xi_0$- and $\xi_1$-gauges that correspond to adding the brane-localized terms,
\begin{gather}
S_0^{\mathrm{GF}}=-\int d^4 x d z \; \delta(z-z_0)\frac{1}{2\xi_0}\Big[\partial_{\mu}A^{\mu}+\xi_0\left(f_0^2 \theta_0+\frac{A_5}{zg^2k}\right)\Big]^2
\end{gather}
and
\begin{gather}
S_1^{\mathrm{GF}}=-\int d^4 x d z \; \delta(z-z_1)\frac{1}{2\xi_1}\Big[\partial_{\mu}A^{\mu}+\xi_1\left(f_1^2 \theta_1-\frac{A_5}{zg^2k}\right)\Big]^2,
\end{gather}
respectively. These suffice to fix the gauge completely. Requiring that the variation of the action vanishes in the bulk gives the bulk equations of motion for the $A^{\mu}$ and 
$A_5$ zero modes,
\begin{align}
z\partial_5\left(z^{-1}\partial_5A^{\mu}\right)&=0,\\
\partial_5\left[z\partial_5\left(\frac{A_5}{z}\right)\right]&=0.
\end{align}
The solutions of the bulk equations of motion are,
\begin{align}
\label{busol1}
A^{\mu}&=B^{\mu}+z^2C^{\mu},\\
\label{busol2}
A_5&=Dz\log{z}+Ez,
\end{align}
where $B^{\mu}$, $C^{\mu}$, $D$ and $E$ do not depend on $z$.  

Requiring that the variation of the action vanishes on the branes gives the conditions,
\begin{align}
\label{amubc} \left(f_{0,1}^2A^{\mu}\pm\frac{1}{g^2zk}\partial_5A^{\mu}\right)\delta A_{\mu}\Big|_{z_{0,1}}&=0,\\
\left(z\xi\partial_5\left(A_5/z\right)\pm\xi_{0,1}\left[f_{0,1}^2\theta_{0,1}-\frac{A_5}{zg^2k}\right]\right)\delta A_5\Big|_{z_{0,1}}&=0,\\
\left(f_{0,1}^2\theta_{0,1}-\frac{A_5}{zg^2k}\right)\delta\theta_{0,1}\Big|_{z_{0,1}}&=0,
\end{align}
where the $+$ and $-$ are for $z=z_0$ and $z=z_1$, respectively.

Substituting the solutions for $A^{\mu}$ and $A_5$ into the boundary variations we find that the zero modes are,
\begin{gather}
A_5=Ez, \hspace{5mm} \theta_0=\frac{E}{f_0^2g^2k}  \hspace{2mm} \textrm{and} \hspace{2mm} \theta_1=\frac{E}{f_1^2g^2k} ,
\end{gather}
where $E$ is unconstrained until fixed by the normalization of the zero mode. We see that we have a scalar zero mode which is partly $A_5$, partly $\theta_0$ and partly 
$\theta_1$. There is no vector zero mode.

Needless to say, the spectrum of zero modes is a physical and gauge-independent quantity,
though its description in terms of fields is not.
Let us, then, reassure ourselves that computing the spectrum in another gauge will give the same result.
Most interesting among these is a gauge in which the boundary scalars $\theta_{0,1}$ vanish, because in this gauge
we recover the original description of the anomalous gauge theory. Now the gauge-fixed action, $S= S_{\mathrm{bulk}}+ S_0 +S_1$,
has contributions
\begin{align}
S_{\mathrm{bulk}} &= \int d^4 x d z \; \Big[-\frac{1}{4(kz)^5g^2}F_{MN}F^{MN} -\frac{1}{2\xi kzg^2}\Big[\partial_{\mu}A^{\mu}+\xi z \partial_5\left(\frac{A_5}{z}\right)\Big] 
+\dots\Big], \nonumber \\
\label{thet0}
S_0 &= \int d^4 x d z \;\delta(z-z_0)\Big[-\frac{1}{4g_0^2}F_{\mu \nu}F^{\mu \nu} + \frac{f_0^2}{2}A_{\mu}A^{\mu} + \dots \Big],\nonumber \\
S_1 &= \int d^4 x d z \;\delta(z-z_1)\Big[-\frac{1}{4g_1^2}F_{\mu \nu}F^{\mu \nu} + \frac{f_1^2}{2}A_{\mu}A^{\mu} + \dots \Big].
\end{align}
The bulk equations of motion for $A^{\mu}$ and $A_5$ are identical to the previous case and so the solutions are those given in (\ref{busol1}) and (\ref{busol2}). The boundary 
variations for $A^{\mu}$ are identical to those shown in (\ref{amubc}) and consequently there is no zero mode for $A^{\mu}$. The boundary variations for $A_5$ are now given 
by
\begin{gather}
\partial_5\left(A_5/z\right)\delta A_5|_{z_{0,1}}=0.
\end{gather}
Substituting the bulk solution for $A_5$ into the above we find the zero mode
\begin{gather}
A_5=Ez,
\end{gather}
where $E$ is unconstrained. So again we find a scalar zero mode and no zero mode for $A^{\mu}$.

Yet another gauge is $A_5 =0$. In this gauge, the residual gauge transformations, 
$\Lambda (x,z)$ such that $\partial_5 \Lambda (x,z) =0$, contain only one arbitrary function of $x$,
so the gauge is fixed completely by a further gauge fixing on just one brane, say at $z_0$.
So now the gauge-fixed action, $S=S_{\mathrm{bulk}}+ S_0 +S_1$, can be chosen to consist of
\begin{align}
S_{\mathrm{bulk}} &= \int d^4 x d z \; \frac{1}{kzg^2}\Big[-\frac{1}{4}F_{\mu\nu}F^{\mu\nu} -\frac{1}{2}\partial_5A_{\mu}\partial_5A^{\mu}+ \dots\Big], \nonumber \\
S_0 &= \int d^4 x d z \;\delta(z-z_0)\Big[-\frac{1}{4g_0^2}F_{\mu \nu}F^{\mu \nu} + \frac{f_0^2}{2}(\partial_{\mu} 
\theta_0-A_{\mu})^2-\frac{1}{2\xi_0}\Big[\partial_{\mu}A^{\mu}+\xi_0f_0^2\theta_0\Big]^2+\dots \Big],
\nonumber \\
\label{a50}
S_1 &= \int d^4 x d z \;\delta(z-z_1)\Big[-\frac{1}{4g_1^2}F_{\mu \nu}F^{\mu \nu} + \frac{f_1^2}{2}(\partial_{\mu} \theta_1-A_{\mu})^2 + \dots \Big].
\end{align}

In this gauge, it is useful to split the gauge field into transverse and longitudinal components defined in $d=4$ momentum space by
\begin{align}
A_{\mu}^T=(\eta^{\nu}_{\mu}-\frac{p_{\mu}p^{\nu}}{p^2})A_{\nu},\nonumber \\
A_{\mu}^L=\frac{p_{\mu}p^{\nu}}{p^2}A_{\nu}.
\end{align}

The bulk equations of motion for the zero modes are
\begin{gather}
z\partial_5\left(z^{-1}\partial_5A^{T,L}_{\mu}\right)=0,
\end{gather}
with solutions
\begin{align}
A^{T\mu}&=B^{\mu}+z^2C^{\mu},\\
A^{L\mu}&=D^{\mu}+z^2E^{\mu}.
\end{align}

The boundary variations in momentum space in this gauge are given by
\begin{align}
\left(f_{0,1}^2A^{T\mu}\pm\frac{1}{g^2zk}\partial_5A^{T\mu}\right)\delta A^T_{\mu}\Big|_{z_{0,1}}&=0,\\
\left(f_0^2A^{L\mu}+\frac{1}{g^2zk}\partial_5A^{L\mu}\right)\delta A^L_{\mu}\Big|_{z_{0}}&=0,\\
\left(f_1^2A^{L\mu}+\frac{1}{g^2zk}\partial_5A^{L\mu}-f^2_1p^{\mu}\theta_1\right)\delta A^L_{\mu}\Big|_{z_{1}}&=0,\\
\theta_0\delta\theta_{0}\Big|_{z_{0}}&=0,\\
p^\mu A^{L}_{\mu}\delta\theta_{1}\Big|_{z_{1}}&=0.
\end{align}
Substituting the bulk solutions for $A^{T\mu}$ and $A^{L\mu}$ into the above, we find that there is no zero mode for $A^{T\mu}$, but there is a zero mode which is part 
$A^{L\mu}$ and part $\theta_1$. The  $A^{L\mu}$ part is, however, unphysical since it does not couple to conserved currents. We are thus left with the scalar zero mode, 
$\theta_1$.

We see, exhaustively, that the spectrum of zero modes is the same in a variety of gauges, and, in particular, in the gauge
in which the original description of the anomalous theory is recovered. The important point in this gauge is that a consistent
EFT treatment requires that we include mass terms for the gauge fields on the boundary (cf. (\ref{thet0})). In retrospect, this hardly seems surprising, given
that the gauge symmetry that would have forbidden such terms on the branes is anomalous. (If we had chosen, perversely,
not to include the mass terms at tree-level, they would be generated nevertheless by loop effects.)

Let us compare with the spectrum of massless modes we would have obtained with vanishing anomalies on both branes, {\em i.e.}
with $\alpha_0 = \alpha_1 =0$. Now gauge invariance is achieved without the boundary scalars. The simplest way to
find the spectrum is to choose $A_5=0$ gauge. The action in this case is
\begin{align}
S&= \int d^4 x d z \; \frac{1}{kzg^2}\Big[-\frac{1}{4}F_{\mu\nu}F^{\mu\nu} -\frac{1}{2}\partial_5A_{\mu}\partial_5A^{\mu}+\dots\Big] \nonumber \\
&-\int d^4 x d z \big[\;\delta(z-z_0)\frac{1}{4g_0^2}F_{\mu \nu}F^{\mu \nu} +\;\delta(z-z_1)\frac{1}{4g_1^2}F_{\mu \nu}F^{\mu \nu} \big].
\end{align}
The bulk equation of motion and boundary variations are
\begin{gather}
z\partial_5\left(z^{-1}\partial_5A^{\mu}\right)=0,
\end{gather}
and
\begin{gather}
\partial_5A^{\mu}\delta A_{\mu}|_{z_0,z_1}=0,
\end{gather}
respectively. We see that now there is a massless gauge boson in the spectrum given by $A^{\mu}=B^{\mu}$, where $B^{\mu}$ is an undetermined constant in $z$.
Similarly, we may consider the case where just one of the brane-localized anomalies is non-vanishing.
The spectrum contains neither vector nor scalar zero modes.

Let us now compare our findings with those obtained previously in the string theory context.
By being careful not to change the degrees of freedom content of the theory, we have found that the zero-mode
spectrum depends solely on whether or not the brane-localized anomaly coefficients, $\alpha_0$ and $\alpha_1$, are vanishing.
It does not depend on their magnitude or relative sign. This is clear from the discussion surrounding Eqs. (\ref{first} - \ref{second}):
$\alpha_0$ and $\alpha_1$ appear in the WZW interaction terms, but not in the quadratic action; the latter determines the spectrum.
To recap, if both $\alpha_0$ and $\alpha_1$ vanish, we get a vector zero mode and if both do not vanish, we get a scalar zero mode.
Otherwise, there are no zero modes.

In the previous literature, it was found (in the case that both anomalies are non-vanishing) that cancelling anomalies by brane-localized twisted Ramond-Ramond states
\cite{Ibanez:1998qp} gave precisely this pattern, whereas cancelling them by bulk Green-Schwarz two-forms \cite{GrootNibbelink:2003gb} gave 
no zero mode for $\alpha_0 = -\alpha_1$, and both vector and scalar zero modes otherwise.
Our explanation for this is as follows. Adding brane-localized states resurrects gauge-invariance on the branes without changing
the degree of freedom content of the theory (the brane-localized states can be gauged away). Adding bulk two-forms,
however, resurrects gauge-invariance on the branes, but also changes the degrees of freedom of the theory.
Indeed, it is clearly impossible to gauge away a bulk field with the resurrected brane-localized gauge symmetry.
In a sense then, our results are just an effective field theorist's vulgarization of the string theory results.

Let us now make two immediate generalizations. Firstly, we consider the case where the gauge symmetry is assumed broken (or rather, non-linearly realized)
{\em ab initio} on one or both branes. The easiest way to realize this in the interval approach is to include
a boundary scalar or scalars by hand \cite{Csaki:2005vy}, just as one does for an anomaly. Of course, now there is no term of the form (\ref{restore}),
but as we have seen, this is irrelevant as far as determining the spectrum is concerned: it is the mere presence
of the scalar (forced upon us in the anomalous case) that changes the spectrum.

Secondly, we can consider what happens in the case of a general non-Abelian bulk gauge group $G$
broken to subgroups $H_{0,1}$ on the respective branes. As in the last paragraph, 
this is easily achieved in the interval approach via boundary scalars.
Yet again, if any of the generators of the subgroups  $H_{0,1}$ has a brane-localized anomaly, 
we should add a boundary scalar for that 
generator on the relevant brane, together with a non-Abelian Wess-Zumino-Witten (WZW) term \cite{Witten:1983tw}, generalizing (\ref{restore}), to cancel the anomaly.
The type of zero mode, if any, corresponding to that generator is then determined
by exactly the same considerations as in the $U(1)$ case above. So, for example, if a would-be massless gauge boson 
(corresponding to a generator in $H_0 \cap H_1$) has an anomaly on one or both branes, we will find in its stead either no
massless state, or a light scalar.

We remark that, although such scalars appear massless at tree-level, they will acquire masses via quantum loops
of propagators stretching from one brane to the other. The masses arise because one brane does not respect
the symmetry group of the other. Thus, boundary scalars, which are Goldstone bosons of the subgroup on
the relevant brane, are really pseudo-Goldstone bosons. The resulting masses, which are non-local in origin,
are finite, and are of order $1/z_1$ in magnitude, further suppressed by a loop factor of order $4\pi$ \cite{Contino:2003ve}.
It is for this reason that we refer to them as light scalars, rather than massless scalars.

Finally, we remind the reader that, although we have focused our attention on theories
on a warped interval (which admit a holographic dual), the anomaly considerations
discussed here apply equally to a theory on an interval with arbitrary geometry.
This is because the brane-localized anomaly structure depends only on the topology of the interval
and the associated fermionic boundary conditions. If the
brane-localized anomalies are non-vanishing, the construction of a consistent EFT is easily done, 
by adding boundary scalars in the same fashion, and the same conclusion 
applies: the spectrum of $d=4$ zero modes in the presence of a brane-localized anomaly
is altered, in that there are either extra, light scalars, or fewer massless gauge bosons.
\section{\label{d=4}Anomalies in d=4}
We saw in the preceding section how a consistent EFT description for an anomalous gauge theory on an interval
in $d=5$, achieved by resurrecting gauge invariance everywhere on the interval, implies extra scalar states in the theory.
These scalar states are either present as light, physical states in the low-energy spectrum in $d=4$, or are eaten
by would-be massless gauge bosons, removing the gauge bosons from the low-energy spectrum in $d=4$.

In this section, we explain the r\^{o}le these scalar states play in the context 
of the AdS/CFT correspondence for a warped geometry:
they ensure 't Hooft matching of anomalies of global symmetries in the $d=4$ dual, at all energy scales.

Before seeing how this comes about, we remind the reader of 't Hooft's argument \cite{tHooft:1980xb}.
Consider a theory with some global symmetry group $G$ that has an anomaly at some energy scale,
meaning that some correlation function of three global currents has non-vanishing divergence, or
equivalently that the divergence of the current is non-vanishing in the presence of a background gauge field.
(In typical examples, the theory is taken to be weakly-coupled at the given scale, 
such that the anomaly is calculable.)
Now add spectator fermions, transforming in representations of $G$ so as to cancel the global $G$ anomalies,
and weakly-gauge the symmetry $G$, with gauge coupling strength $g \ll 1$. 
As we run down to a lower energy scale, the gauge theory we have constructed remains, of course, anomaly-free. 
If we then further take the limit in which $g\rightarrow 0$,
decoupling the gauge fields, 
we again end up with a theory with anomaly-free global symmetry group $G$.
This theory still contains the decoupled spectator fermions (since they were only ever weakly coupled
to other sectors), whose contribution to the global $G$ anomalies is the same as it was at the higher scale.
This implies that the global $G$ anomalies of the original theory (without spectator fermions or gauge fields)
cannot have changed either. That is, the global anomalies of the original theory must match at all energy scales. 
This holds true even though the theory may have gone through one or more strong-coupling transitions, such that
the effective weakly-coupled degrees of freedom (if any) may be completely changed. 
The weakly-coupled degrees of freedom that contribute to the anomaly include fermions 
transforming in representations of $G$, together with scalar fields, which can contribute
to the anomaly via Wess-Zumino-Witten (WZW) terms in the effective action \cite{Witten:1983tw}.

The $d=4$ holographic dual of an $AdS_5$ theory seems to lend itself ideally to a study of 't Hooft matching, in
that the $z$ co-ordinate (more precisely, its logarithm) corresponds to the energy scale in the $d=4$ CFT. 
This is because the combination of a constant scale transformation $x^{\mu}\rightarrow a x^{\mu}$ 
in the co-ordinates  of the $d=4$ dual, accompanied by a scaling $z \rightarrow a z$,
amounts to an isometry of the $AdS_5$ metric (\ref{adsmetric}), and so does not change the physics.
This is what we expect for a theory with conformal symmetry, provided we interpret $\log z$ as the energy scale.
Thus we expect, intuitively, that the non-trivial $z$ dependence of the anomalies in the $d=5$ theory encodes
information about the anomalies at different energy scales in the $d=4$ theory, which
is precisely the context in which the 't Hooft anomaly matching condition is applied.

To set the scene,
let us review Witten's original observation \cite{Witten:1998qj}, relating the anomalies in the $d=4$ theory and the Chern-Simons term
in an AdS space without branes. We start from Witten's conjecture for the AdS/CFT correspondence in the form
\begin{gather} \label{CFT}
\langle \exp{\int d^4 x \; \mathcal{O}\varphi}\rangle_{\mathrm{CFT}} = \int_{\Phi(z\rightarrow 0)\rightarrow \varphi}\hspace{-1.4cm} D \Phi \;
\exp {iS[\Phi]},
\end{gather}
where the right-hand side represents a path-integral in AdS with respect to generic bulk field $\Phi (x,z)$, with
the restriction that $\Phi (x,z)$ tends to the value $\varphi (x)$ on the projective boundary of AdS, 
given by $z \rightarrow 0$.
According to the conjecture, this is equivalent to the correlation function of a CFT deformed by $\int \mathcal{O \varphi}$,
where $\mathcal{O}$ is some operator of the CFT dual to the source $\varphi (x)$. 
In the limit in which we are interested, where the CFT has large rank and strong ('t Hooft) coupling, 
the path integral on the right-hand side
is understood to be computed on-shell, {\em i.e.} the action is evaluated subject to the condition that the 
classical equations of motion are satisfied.
As it stands, the correspondence is ill-defined: the action on the right-hand side
has a divergence that comes from integrating the Lagrangian density over the IR of AdS (the region of small $z$).
If the correspondence has any chance of being true, it must be that the left-hand side is also divergent, and
indeed it is: the divergence is a UV divergence arising because we have deformed the
CFT. This corroborates our previous claim that $\log z$ corresponds 
to the energy scale in the $d=4$ dual: the region of small $z$ in $AdS_5$ corresponds to the UV region of the CFT.
More specifically, we can remove the divergence on the AdS side, by truncating $AdS_5$ at some small value $z=z_0$;
this truncation must correspond to cutting-off the deformed CFT in the UV in some 
definite fashion at a scale $\sim 1/z_0$.
The conjectured correspondence is thus modified to
\begin{gather} \label{CFT'}
\langle \exp{\int d^4 x \; \mathcal{O}\varphi}\rangle_{\mathrm{CFT'}} = \int_{\Phi(x, z_0)= \varphi (x)} \hspace{-1.65cm}D \Phi \;
\exp{ iS[\Phi]},
\end{gather}
where we have put a prime on the left-hand side to indicate that the deformed CFT has been cut-off in the UV.
We note in passing, that cutting-off the CFT in the UV will, in general, induce kinetic terms for the generic source field 
$\varphi (x)$ (a source is, from the EFT point of view, just a higher-dimensional field), so we are free
to make $\varphi (x)$ a dynamical field if we so choose, by performing a path-integral with respect to it on both sides
of (\ref{CFT'}).

In the particular case of a bulk gauge field,
$A_M (x,z)$, we might write the correspondence as
\begin{gather} \label{ACFT}
\langle \exp{\int d^4 x \; J_{\mu} a^{\mu}}\rangle_{\mathrm{CFT'}} =\hspace{-0.7cm} \int \limits_{\begin{subarray}{l} A_{\mu}(x, z_0) = a_{\mu}(x)\\ A_5(x, z_0)= 
0\end{subarray}} 
\hspace{-0.9cm}D A_M \;
\exp {iS[A_M]}.
\end{gather}
For the time being, we choose not to path-integrate with respect to the boundary field $a_{\mu}(x)=A_{\mu}(x, z_0)$, meaning this is a background
gauge field in the $d=4$ dual.
Now, assuming that the $d=5$ integrand is gauge-invariant, we can show (at least formally) that the global CFT current $J_{\mu}$ is 
conserved. Indeed, consider making a change of variables in the $d=5$ theory that takes
the form of a bulk gauge transformation, $A_M \rightarrow A_M + \partial_M \Lambda (x,z)$, with the restriction that $\partial_5 \Lambda (x,z_0)=0$.
Since the integrand is gauge-invariant,
we find 
\begin{gather}
\int\limits_{\begin{subarray}{l} A_{\mu}(x, z_0) = a_{\mu}(x)\\ A_5(x, z_0)= 0\end{subarray}} 
\hspace{-0.9cm}D A_M \;
\exp {iS[A_M]} =\hspace{-1.3cm} \int\limits_{\begin{subarray}{l} A_{\mu}(x, z_0) = a_{\mu}(x)+\partial_{\mu}\lambda (x)\\ A_5(x, z_0)= 0\end{subarray}} 
\hspace{-1.45cm}D A_M \;
\exp {iS[A_M]},
\end{gather}
where $\Lambda (x,z_0) = \lambda (x)$, implying
\begin{gather}
\frac{\delta}{\delta \lambda (x)}\langle \exp{\int d^4 x \; J_{\mu} (a^{\mu}+ \partial^{\mu}\lambda)}\rangle_{\mathrm{CFT'}} = 
0.
\end{gather}
Thus
\begin{gather}
\langle \partial_{\mu}J^{\mu}e^{\int d^4 x \; J_{\mu} a^{\mu}} \rangle_{\mathrm{CFT'}}=0.
\end{gather}
Going further, we may ask what happens if the $d=5$ action contains a Chern-Simons term, such that the action
is gauge-invariant everywhere, except on the $z_0$ boundary. For a Chern-Simons term of the form (\ref{CS}), the same
argument implies that
\begin{gather}
\langle (\partial^{\mu}J^{\mu}+cf_{\mu \nu}\tilde{f}^{\mu \nu}) e^{\int d^4 x \; J_{\mu} a^{\mu}} \rangle_{\mathrm{CFT'}}=0,
\end{gather}
where $f_{\mu \nu} = \partial_{\mu}a_{\nu} - \partial_{\nu} a_{\mu}$.
So we find that the bulk Chern-Simons term produces an anomaly in the global current of the cut-off, deformed CFT,
in the presence of the background gauge field $a^{\mu}$.
In particular, in the case of Type IIB supergravity on $AdS_5$, one finds that the Chern-Simons term for
$SU(4)$ bulk gauge fields reproduces the global anomaly of the $SU(4)$ $R$-symmetry currents in the $d=4$ dual,
which is $\mathcal{N}=4$ supersymmetric Yang-Mills theory \cite{Witten:1998qj, Freedman:1998tz}.

On reflection, the argument we have just reviewed is suspect, because the path integral on the right-hand side 
of (\ref{ACFT}) involves integration over infinitely many, physically-equivalent, gauge field configurations. 
In order to make sense of this path-integral, we need to regulate it by, say, the procedure
of Fadeev and Popov, or more generally, BRST. But such a gauge-fixing procedure invalidates the argument just given,
which invoked gauge-invariance of the integrand in the $d=5$ theory.

We can recover the argument in a refined form by regulating the $d=5$ path integral in such a way that
it is rendered finite (at least at tree-level), but such that there remains a residual bulk gauge symmetry to
which a background gauge transformation $\lambda(x)$ can be smoothly lifted.
A suitable regulator is provided by the delta-functional $\delta(A_5(x,z))$ (for which the Faddev-Popov determinant
is trivial), whose argument is invariant under the residual gauge transformations $\Lambda (x,z)$
such that  $\partial_5 \Lambda (x,z) =0$, {\em i.e.} those for which $\Lambda (x,z)$ is independent of $z$.
We note that this is compatible with the boundary condition, $A_5 (x,z_0)=0$, chosen on the $z_0$ brane.
Furthermore, with this BC, the regulator $\delta(A_5(x,z))$ can be used even when the theory is not gauge-invariant on the $z_0$ brane,
because of an anomaly.
This is because we have chosen $A_5 =0$ as a boundary condition there, for which
the argument of the delta-functional vanishes identically.
With the delta-functional regulator included, the correspondence becomes
\begin{gather} \label{ACFTR}
\langle \exp{\int d^4 x \; J_{\mu} a^{\mu}}\rangle_{\mathrm{CFT'}} =\hspace{-0.5cm} \int \limits_{\begin{subarray}{l} A_{\mu}(x, z_0) = a_{\mu}(x)\\ A_5(x, z_0)= 
0\end{subarray}} 
\hspace{-0.9cm}D A_M \; \delta(A_5(x,z))
\exp {iS[A_M]},
\end{gather}
and the path integral on the right-hand side is rendered finite in such a way that the argument given previously
still goes through. We shall make frequent use of this argument,
which we call the holographic anomaly argument,
in the sequel. We repeat that it can be employed whenever the $d=5$ theory is gauge-invariant for $z>z_0$. 

Hitherto, we have only considered the possibility of gauge-variance on the $z_0$ brane resulting from Chern-Simons terms.
If we have a bulk fermion $\Psi$ charged under the gauge group, then it should also be present in the path integral on
the right-hand side, including the path-integration with respect to its value on the $z_0$-brane.
With the types of fermionic BCs we have considered, only one Weyl component of $\Psi$, $\psi$ say, can be non-vanishing
and (\ref{ACFTR}) is modified to
\begin{gather} \nonumber
\int D \psi D\overline{\psi}\; e^{iS_0[\psi,a^{\mu}]}\langle e^{\int d^4 x \; 
J_{\mu} a^{\mu}+\mathcal{O}_{\psi}\psi}\rangle_{\mathrm{CFT'}} = \int D \psi D\overline{\psi} \; e^{iS_0[\psi,a^{\mu}]}\hspace{-0.8cm} \int\limits_{\begin{subarray}{l} 
A_{\mu}(x, z_0) = a_{\mu}(x)\\ A_5(x, z_0)= 0\\ \Psi (x,z_0)=\psi (x)\end{subarray}} 
\hspace{-0.9cm} D A_M D \Psi D\overline{\Psi} \;
e^{iS[A_M,\Psi]},
\end{gather}
where we included a possible brane-localized action, $S_0$, for the fermion.
When we invoke the holographic anomaly argument, we will pick up an extra contribution  to
gauge-variance on the $z_0$ brane coming from the Jacobean of the bulk fermion measure under the gauge transformation. 
This Jacobean is determined directly from (\ref{bulkanom}), where, because there is currently only a single brane (at $z=z_0$),
we discard the $z_1$ piece.
The holographic anomaly argument then gives
\begin{gather}\label{4anom}
\langle (\partial^{\mu}J^{\mu}+cf_{\mu \nu}\tilde{f}^{\mu \nu}) e^{\int d^4 x \; J_{\mu} a^{\mu}} \rangle_{\mathrm{CFT'}}=0,
\end{gather}
where now $c$ contains contributions from both the Chern-Simons terms and the bulk fermion.

Similarly, for a dynamical brane-localized fermion $\psi'$ at $z=z_0$, we should take (\ref{ACFTR}), include the brane-localized action
for the fermion, and path-integrate with respect to it, obtaining
\begin{gather} 
\int D \psi' D\overline{\psi'}\; e^{iS_0[\psi,a^{\mu}]}\langle e^{\int d^4 x \; J_{\mu} a^{\mu}}\rangle_{\mathrm{CFT'}} =\hspace{-0.6cm} \int\limits_{\begin{subarray}{l} 
A_{\mu}(x, z_0) = a_{\mu}(x)\\ A_5(x, z_0)= 0\end{subarray}} 
\hspace{-0.9cm}D A_M D \psi' D\overline{\psi'} \;
e^{iS_0[\psi',a^{\mu}]}e^{iS[A_M]}.
\end{gather}
Now when we make the holographic anomaly argument, we find an expression like (\ref{4anom}), but now the coefficient $c$ contains 
contributions from Chern-Simons terms and brane-localized fermions. 

In this way, we see how the $z_0$-brane-localized gauge anomaly of the $d=5$ theory computes, via
the holographic anomaly argument, the global anomaly of the $d=4$ dual, which is a strongly-coupled CFT coupled to
external fields, in a background gauge field. This anomaly is, of course, the anomaly computed at the scale $1/z_0$.

To see how 't Hooft matching works in the $d=4$ dual, we need to compute the anomaly at some lower energy
scale, $1/z'$,say. To do so, we need to compute the EFT for the $d=4$ dual obtained by integrating out the physics
at energy scales corresponding to $z_0<z<z'$. In the $d=5$ picture, we do this by shifting the position of the 
$z_0$ brane to $z=z'$, whilst demanding that the physics at $z>z'$ remains the same. At 
tree-level, this amounts to modifying the brane-localized action at $z'$ such that the solutions of the $d=5$ equations of motion remain the same for $z>z'$.
 The details 
 of this `holographic renormalization group flow' are
 described in \cite{Verlinde:1999fy,Lewandowski:2002rf}; as far as the anomalies are concerned though, the flow is trivial. The theory with brane at $z'$ 
 contains the same field content as the theory with brane at $z_0$, and therefore the brane-localized anomalies remain the same.
As a corollary, the global anomalies of the $d=4$ dual cut-off at $1/z'$ are the same for all $z'>z_0$,
and 't Hooft matching is trivial.

So far, we have explicitly excluded the brane at $z=z_1$ from our arguments.
Re-instating the brane corresponds \cite{Arkani-Hamed:2000ds, Rattazzi:2000hs} to spontaneously breaking the conformal symmetry in the IR at energy scale $1/z_1$.
At energy scales lower than $1/z_1$, the correspondence no longer holds; the $d=4$ theory must be studied in the normal
way, without recourse to holographic arguments. 
We know from the previous section that adding the brane at $z=z_1$ introduces further gauge anomalies into the $d=5$ theory.
What, then, do these correspond to on the dual $d=4$ side? In trying to generalize the arguments we have just 
made for the theory with only a single brane at $z=z_0$, we immediately encounter a problem: the presence
of a non-vanishing anomaly on the $z_1$ brane prohibits us {\em a priori} from making the holographic anomaly argument,
because this assumed gauge invariance for $z>z_0$; the existence of the anomaly implies that there is no gauge symmetry at $z=z_1$.
Unsurprisingly, the solution is to resurrect the gauge symmetry at $z=z_1$, by adding a scalar field $\theta_1$
together with a brane-localized term of the form (\ref{restore}), just as we did before. We stress again that we are not
changing the theory or its degrees of freedom, but merely furnishing ourselves with an equivalent description.

The beauty of this equivalent description, with resurrected gauge symmetry on the $z_1$ brane,
is that we can invoke the holographic anomaly argument once more.\footnote{Note that
we have not bothered to resurrect the gauge symmetry on the $z_0$ brane, since the holographic anomaly argument
is not contingent upon it.} When we do so, we find that the global anomaly of the $d=4$ dual at the scale 
$1/z_0$ is computed by the gauge anomaly on the $z_0$ brane in the $d=5$ theory, and that by shifting the position of the
$z_0$ brane to $z_0 \leq z'<z_1$ whilst keeping the physics the same, the global anomaly trivially obeys the 't Hooft matching condition
at all scales $1/z_0 \geq 1/z'>1/z_1$.
What happens at $z'=z_1$? In the $d=5$ picture, this corresponds to the endpoint of the holographic RG flow, 
with the cut-off brane at $z'$ hitting the IR brane at $z_1$. Now the entire bulk has been integrated out, and 
we are left with a $d=4$ EFT with cut-off $1/z_1$, obtained by adding the 
$z'$-brane-localized action (with $z' \rightarrow z_1$)
to the $z_1$-brane-localized action \cite{Lewandowski:2002rf}. If we make the holographic anomaly argument at $z'=z_1$, we will find that the global
anomaly of the $d=4$ dual cut-off at $z_1$ is still the same as the anomaly at $z'<z_1$: contributions to the $z_1$-brane-localized anomaly, 
whether they come from Chern-Simons terms, or bulk or brane-localized fermions, are cancelled by the boundary
scalar $\theta_1$, which appears as a light state in the low-energy $d=4$ theory. 
What happens at energy scales below $1/z_1$? Now the correspondence breaks down, but we can still compute the effective action
by taking the effective theory with cut-off $1/z_1$ and integrating down to the lower energy scale in the normal way. The global
anomalies must still match, and, provided the degrees of freedom remain weakly-coupled, come from
the light fermions and scalars.

We remind the reader that, so far in this section we have kept the gauge field $a^{\mu}(x)$ 
as a non-dynamical, background gauge field, since this is what we need for computing the global anomalies of the
$d=4$ dual theory. We can also choose to make the gauge field dynamical, by 
path-integrating with respect to it on both sides of, {\em e.g.} (\ref{ACFTR}). The 
fate of a light scalar (assuming there is a non-vanishing anomaly for some generator on the $z_1$ brane) then
 depends on whether or not there is an anomaly for the generator on the $z_0$ brane. If there is an anomaly on the $z_0$ brane,
then a mass term will be generated for the gauge field in the $d=4$ dual EFT at the scale $1/z_0$; it will not be present in the low-energy spectrum, but
the scalar will. If, by contrast, there is no anomaly on the $z_0$ brane, the dynamical gauge field will
be present as a massless gauge field in the theory cut-off at energy scales above $1/z_1$. At the energy scale
$1/z_1$, it will `eat' the scalar $\theta_1$, leaving neither a gauge boson nor a scalar in the low-energy
spectrum. These spectra are, of course, just those we determined in the previous section.

What would happen if we chose not to resurrect the gauge symmetry on the $z_1$ brane by adding the scalar $\theta_1$?
Now we cannot use the holographic anomaly argument to compute the global anomalies of the $d=4$ dual, 
because the theory is not gauge invariant for all $z>z_0$. That said, the spectrum of $d=4$ zero modes,
including possibly a light scalar degree of freedom, cannot change, since the spectrum cannot depend on
which equivalent description of the theory we choose. Reassuringly, the spectrum does not change:
as we saw in the previous section, the scalar is provided in this description by a zero mode of the $A_5$ field,
rather than by $\theta_1$. What does change is that we cannot compute the global anomaly of the $d=4$ dual in this 
description, because we simply do not have a theorem, in the form of the holographic anomaly argument, available to us.
\section{\label{phen}Implications for Warped Models of EWSB}
We have tacitly assumed a $U(1)$ bulk gauge symmetry in the foregoing, but 
generalization to the case where bulk gauge group $G$ is broken to subgroups $H_{0,1}$ on the respective branes
is straightforward. We simply analyse each generator of the bulk Lie algebra in turn. If the generator
is assumed to be unbroken on a given brane ({\em i.e.} contained in $H_0$ or $H_1$, as appropriate)
then we should consider whether there is an anomaly, involving that generator, localized on the given brane, due to any of the sources
we discussed above. If there is an anomaly, then we can find a consistent EFT by adding a boundary scalar to
restore gauge invariance. This means, though, that just as in the U(1) case, the spectrum of
massless states in $d=4$ will be changed, with either extra light scalars or fewer massless gauge bosons. 
So, in general, all of the $H_{0,1,}$ anomalies on the respective branes must be cancelled in order that the spectrum
of massless states be that which is assumed {\em a priori}.

This observation is of some relevance for warped models of electroweak symmetry-breaking in the Standard Model (SM).

Take, as an example, the Composite Higgs Model of \cite{Contino:2006qr}. This model employs a bulk gauge group $G=SU(3)_c\times SO(5)\times U(1)_X$, which is broken 
down to $H_1=SU(3)_c\times O(4)\times U(1)_X$ on the IR boundary, and $H_0=SU(3)_c\times SU(2)_L\times U(1)_Y$ on the UV boundary, where hypercharge is defined as 
$Y=X+T_3^R$.  The Higgs appears as a pseudo-Goldstone Boson consisting of the four real scalar fields that corresponds to the $A_5$ components of the $SO(5)/SO(4)$ 5D gauge 
fields. There are various ways to obtain the fermion content of the SM. In \cite{Contino:2006qr}, two possibilities are presented, labeled MCHM$_5$ (where MCHM stands for 
Minimal Composite Higgs Model) and MCHM$_{10}$, where the SM fermions are contained in the ${\bf 5}$ and ${\bf 10}$ representations of $SO(5)$, respectively. 

Let us first examine in detail the MCHM$_{5}$ model. Each SM generation of quarks is identified with zero modes of the following bulk multiplets (where BCs are indicated 
explicitly) \cite{Contino:2006qr} ,
\begin{align}
\nonumber
\xi_{q_1}&=\left[\begin{array}{ll} ({\bf 2,2})^{q_1}_L=\left[\begin{array}{ll} q^{\prime}_{1L}(- +) \\ q_{1L}(+ +)  \end{array}\right] & ({\bf 2,2})^{q_1}_R=\left[\begin{array}{ll} 
q^{\prime}_{1R}(+ -) \\ q_{1R}(- -)  \end{array}\right] \\ (1, 1)^{q_1}_{L}(- -) & (1, 1)^{q_1}_R(+ +) \end{array}\right], \hspace{2mm}
\xi_u=\left[\begin{array}{ll} ({\bf 2,2})^u_L(+ -) & ({\bf 2,2})^u_R(- +) \\ (1, 1)^u_L(- +) & (1, 1)^u_R(+ -)  \end{array}\right] \\
\nonumber\\
\xi_{q_2}&=\left[\begin{array}{ll} ({\bf 2,2})^{q_2}_L=\left[\begin{array}{ll} q_{2L}(+ +) \\ q^{\prime}_{2L}(- +)  \end{array}\right] & ({\bf 2,2})^{q_2}_R=\left[\begin{array}{ll} 
q_{2R}(- -) \\ q^{\prime}_{2R}(+ -)  \end{array}\right] \\ (1, 1)^{q_2}_{L}(- -) & (1, 1)^{q_2}_R(+ +) \end{array}\right], \hspace{2mm}
\xi_d=\left[\begin{array}{ll} ({\bf 2,2})^d_L(+ -) & ({\bf 2,2})^d_R(- +) \\ (1, 1)^d_L(- +) & (1, 1)^d_R(+ -)  \end{array}\right],
\end{align}
where $\xi_{q_1}$, $\xi_u$ ($\xi_{q_2}$, $\xi_d$) transform as ${\bf 5}_{2/3}$ (${\bf 5}_{-1/3}$) multiplets of $SO(5)\times U(1)_X$. The fields are grouped within the $\xi$ 
multiplets in representations of $SO(4)\sim SU(2)_L\times SU(2)_R$, using the decomposition ${\bf 5}={\bf 4}\oplus{\bf 1}=({\bf 2, 2})\oplus ({\bf 1, 1})$.
For each generation there is an additional right-handed (chirality under $4D$ Lorentz group) doublet of $SU(2)_L$, $\tilde{q}_R=({\bf 2, 0})^q_R$, localized on the UV brane. 
These boundary quarks transform as ${\bf 3}$s of $SU(3)$ and have $Y=1/6$. It is suggested in \cite{Contino:2006qr} that the leptons be embedded in a similar manner to the 
quarks but with differing $U(1)_X$ charges. Doing this we have the following bulk multiplets,
\begin{align}
\nonumber
\xi_{l_1}&=\left[\begin{array}{ll} ({\bf 2,2})^{l_1}_L=\left[\begin{array}{ll} l^{\prime}_{1L}(- +) \\ l_{1L}(+ +)  \end{array}\right] & ({\bf 2,2})^{l_1}_R=\left[\begin{array}{ll} 
l^{\prime}_{1R}(+ -) \\ l_{1R}(- -)  \end{array}\right] \\ (1, 1)^{l_1}_{L}(- -) & (1, 1)^{l_1}_R(+ +) \end{array}\right], \hspace{2mm}
\xi_{\nu}=\left[\begin{array}{ll} ({\bf 2,2})^{\nu}_L(+ -) & ({\bf 2,2})^{\nu}_R(- +) \\ (1, 1)^{\nu}_L(- +) & (1, 1)^{\nu}_R(+ -)  \end{array}\right] \\
\nonumber\\
\xi_{l_2}&=\left[\begin{array}{ll} ({\bf 2,2})^{l_2}_L=\left[\begin{array}{ll} l_{2L}(+ +) \\ l^{\prime}_{2L}(- +)  \end{array}\right] & ({\bf 2,2})^{l_2}_R=\left[\begin{array}{ll} 
l_{2R}(- -) \\ l^{\prime}_{2R}(+ -)  \end{array}\right] \\ (1, 1)^{l_2}_{L}(- -) & (1, 1)^{l_2}_R(+ +) \end{array}\right], \hspace{2mm}
\xi_e=\left[\begin{array}{ll} ({\bf 2,2})^e_L(+ -) & ({\bf 2,2})^e_R(- +) \\ (1, 1)^e_L(- +) & (1, 1)^e_R(+ -)  \end{array}\right],
\end{align}
where $\xi_{l_1}$, $\xi_{\nu}$ ($\xi_{l_2}$, $\xi_e$) transform as ${\bf 5}_{0}$ (${\bf 5}_{-1}$) multiplets of $SO(5)\times U(1)_X$. 
In analogy with the quarks, for each generation of leptons there is an additional right-handed doublet of $SU(2)_L$, $\tilde{l}_R=({\bf 2, 0})^l_R$, localized on the UV brane. These 
boundary leptons have $Y=-1/2$.
We find that for the fermion content described above there are no anomalies for either $H_0$ or $H_1$ in the MHCM$_5$. Consequently, this particular version of the MCHM 
generates the correct spectra of SM gauge and Higgs bosons, without recourse to Chern-Simons terms.
We have also computed the anomalies for the MCHM$_{10}$ and various other warped models of SM EWSB, both with and without a Higgs. In many of these models the full 
content of the fermionic sector is somewhat ambiguous. However, in the same way as we have done for the MCHM$_5$ we can write down a plausible fermionic sector and 
calculate the resulting fermionic anomalies. We find that in the majority of these models there are non-vanishing brane-localized fermion anomalies which need to be cancelled in 
order to get the correct description of the SM at low energies.  Two possible ways to cancel these anomalies are either to add Chern-Simons terms to the $d=5$ theory, and/or, more 
generally, to change the fermion content of the model; such changes will affect the low-energy phenomenology, which we hope to explore in future work \cite{gmw}.

\begin{acknowledgments}
We thank J. March-Russell for useful discussions. This work was partially supported by the EU FP6 Marie Curie Research and Training Network ``UniverseNet". 
(HPRN-CT-2006-035863)
\end{acknowledgments}

\begin{thebibliography}{41}
\expandafter\ifx\csname natexlab\endcsname\relax\def\natexlab#1{#1}\fi
\expandafter\ifx\csname bibnamefont\endcsname\relax
  \def\bibnamefont#1{#1}\fi
\expandafter\ifx\csname bibfnamefont\endcsname\relax
  \def\bibfnamefont#1{#1}\fi
\expandafter\ifx\csname citenamefont\endcsname\relax
  \def\citenamefont#1{#1}\fi
\expandafter\ifx\csname url\endcsname\relax
  \def\url#1{\texttt{#1}}\fi
\expandafter\ifx\csname urlprefix\endcsname\relax\def\urlprefix{URL }\fi
\providecommand{\bibinfo}[2]{#2}
\providecommand{\eprint}[2][]{\url{#2}}

\bibitem[{\citenamefont{Maldacena}(1998)}]{Maldacena:1997re}
\bibinfo{author}{\bibfnamefont{J.~M.} \bibnamefont{Maldacena}},
  \bibinfo{journal}{Adv. Theor. Math. Phys.} \textbf{\bibinfo{volume}{2}},
  \bibinfo{pages}{231} (\bibinfo{year}{1998}), \eprint{hep-th/9711200}.

\bibitem[{\citenamefont{Witten}(1998)}]{Witten:1998qj}
\bibinfo{author}{\bibfnamefont{E.}~\bibnamefont{Witten}},
  \bibinfo{journal}{Adv. Theor. Math. Phys.} \textbf{\bibinfo{volume}{2}},
  \bibinfo{pages}{253} (\bibinfo{year}{1998}), \eprint{hep-th/9802150}.

\bibitem[{\citenamefont{Arkani-Hamed
  et~al.}(2001{\natexlab{a}})\citenamefont{Arkani-Hamed, Porrati, and
  Randall}}]{Arkani-Hamed:2000ds}
\bibinfo{author}{\bibfnamefont{N.}~\bibnamefont{Arkani-Hamed}},
  \bibinfo{author}{\bibfnamefont{M.}~\bibnamefont{Porrati}}, \bibnamefont{and}
  \bibinfo{author}{\bibfnamefont{L.}~\bibnamefont{Randall}},
  \bibinfo{journal}{JHEP} \textbf{\bibinfo{volume}{08}}, \bibinfo{pages}{017}
  (\bibinfo{year}{2001}{\natexlab{a}}), \eprint{hep-th/0012148}.

\bibitem[{\citenamefont{Rattazzi and Zaffaroni}(2001)}]{Rattazzi:2000hs}
\bibinfo{author}{\bibfnamefont{R.}~\bibnamefont{Rattazzi}} \bibnamefont{and}
  \bibinfo{author}{\bibfnamefont{A.}~\bibnamefont{Zaffaroni}},
  \bibinfo{journal}{JHEP} \textbf{\bibinfo{volume}{04}}, \bibinfo{pages}{021}
  (\bibinfo{year}{2001}), \eprint{hep-th/0012248}.

\bibitem[{\citenamefont{Perez-Victoria}(2001)}]{Perez-Victoria:2001pa}
\bibinfo{author}{\bibfnamefont{M.}~\bibnamefont{Perez-Victoria}},
  \bibinfo{journal}{JHEP} \textbf{\bibinfo{volume}{05}}, \bibinfo{pages}{064}
  (\bibinfo{year}{2001}), \eprint{hep-th/0105048}.

      \bibitem[{\citenamefont{Callan and Harvey}(1985)}]{Callan:1984sa}
	\bibinfo{author}{\bibfnamefont{L.~D.}~\bibnamefont{Faddeev}} \bibnamefont{and}
  \bibinfo{author}{\bibfnamefont{S.~L.}~\bibnamefont{Shatashvili}},
    \bibinfo{journal}{Theor. Math. Phys} \textbf{\bibinfo{volume}{60}},
	\bibinfo{pages}{770}, (\bibinfo{year}{1985});
\bibinfo{author}{\bibfnamefont{C.~G.}~\bibnamefont{Callan}} \bibnamefont{and}
  \bibinfo{author}{\bibfnamefont{J.~A.}~\bibnamefont{Harvey}},
    \bibinfo{journal}{Nucl. Phys.} \textbf{\bibinfo{volume}{B250}},
  \bibinfo{pages}{427} (\bibinfo{year}{1985}).
  
\bibitem[{\citenamefont{Arkani-Hamed
  et~al.}(2001{\natexlab{b}})\citenamefont{Arkani-Hamed, Cohen, and
  Georgi}}]{Arkani-Hamed:2001is}
\bibinfo{author}{\bibfnamefont{N.}~\bibnamefont{Arkani-Hamed}},
  \bibinfo{author}{\bibfnamefont{A.~G.} \bibnamefont{Cohen}}, \bibnamefont{and}
  \bibinfo{author}{\bibfnamefont{H.}~\bibnamefont{Georgi}},
  \bibinfo{journal}{Phys. Lett.} \textbf{\bibinfo{volume}{B516}},
  \bibinfo{pages}{395} (\bibinfo{year}{2001}{\natexlab{b}}),
  \eprint{hep-th/0103135}.

\bibitem[{\citenamefont{Scrucca et~al.}(2002)\citenamefont{Scrucca, Serone,
  Silvestrini, and Zwirner}}]{Scrucca:2001eb}
\bibinfo{author}{\bibfnamefont{C.~A.} \bibnamefont{Scrucca}},
  \bibinfo{author}{\bibfnamefont{M.}~\bibnamefont{Serone}},
  \bibinfo{author}{\bibfnamefont{L.}~\bibnamefont{Silvestrini}},
  \bibnamefont{and} \bibinfo{author}{\bibfnamefont{F.}~\bibnamefont{Zwirner}},
  \bibinfo{journal}{Phys. Lett.} \textbf{\bibinfo{volume}{B525}},
  \bibinfo{pages}{169} (\bibinfo{year}{2002}), \eprint{hep-th/0110073}.
  


\bibitem[{\citenamefont{Barbieri et~al.}(2002)\citenamefont{Barbieri, Contino,
  Creminelli, Rattazzi, and Scrucca}}]{Barbieri:2002ic}
\bibinfo{author}{\bibfnamefont{R.}~\bibnamefont{Barbieri}},
  \bibinfo{author}{\bibfnamefont{R.}~\bibnamefont{Contino}},
  \bibinfo{author}{\bibfnamefont{P.}~\bibnamefont{Creminelli}},
  \bibinfo{author}{\bibfnamefont{R.}~\bibnamefont{Rattazzi}}, \bibnamefont{and}
  \bibinfo{author}{\bibfnamefont{C.~A.} \bibnamefont{Scrucca}},
  \bibinfo{journal}{Phys. Rev.} \textbf{\bibinfo{volume}{D66}},
  \bibinfo{pages}{024025} (\bibinfo{year}{2002}), \eprint{hep-th/0203039}.

\bibitem[{\citenamefont{von Gersdorff and Quiros}(2003)}]{vonGersdorff:2003dt}
\bibinfo{author}{\bibfnamefont{G.}~\bibnamefont{von Gersdorff}}
  \bibnamefont{and} \bibinfo{author}{\bibfnamefont{M.}~\bibnamefont{Quiros}},
  \bibinfo{journal}{Phys. Rev.} \textbf{\bibinfo{volume}{D68}},
  \bibinfo{pages}{105002} (\bibinfo{year}{2003}), \eprint{hep-th/0305024}.

\bibitem[{\citenamefont{Scrucca and Serone}(2004)}]{Scrucca:2004jn}
\bibinfo{author}{\bibfnamefont{C.~A.} \bibnamefont{Scrucca}} \bibnamefont{and}
  \bibinfo{author}{\bibfnamefont{M.}~\bibnamefont{Serone}},
  \bibinfo{journal}{Int. J. Mod. Phys.} \textbf{\bibinfo{volume}{A19}},
  \bibinfo{pages}{2579} (\bibinfo{year}{2004}), \eprint{hep-th/0403163}.
  
    
   \bibitem[{\citenamefont{Boyarsky et~al}(2005)\citenamefont{ Boyarsky, Ruchayskiy and Shaposhnikov}}]{Boyarsky:2005eq} 
       \bibinfo{author}{\bibfnamefont{A.} \bibnamefont{Boyarsky}},
	  \bibinfo{author}{\bibfnamefont{O.}~\bibnamefont{Ruchayskiy}},  \bibnamefont{and}
  \bibinfo{author}{\bibfnamefont{M.}~\bibnamefont{Shaposhnikov}},
  \bibinfo{journal}{Phys. Rev.} \textbf{\bibinfo{volume}{D72}},
  \bibinfo{pages}{085011} (\bibinfo{year}{2005}), \eprint{arXiv:hep-th/0507098};
    \bibinfo{author}{\bibfnamefont{A.} \bibnamefont{Boyarsky}},
	  \bibinfo{author}{\bibfnamefont{O.}~\bibnamefont{Ruchayskiy}},  \bibnamefont{and}
  \bibinfo{author}{\bibfnamefont{M.}~\bibnamefont{Shaposhnikov}},
  \bibinfo{journal}{Phys. Lett.} \textbf{\bibinfo{volume}{B626}},
  \bibinfo{pages}{184} (\bibinfo{year}{2005}), \eprint{hep-ph/0507195}.


\bibitem[{\citenamefont{Hirayama and Yoshioka}(2004)}]{Hirayama:2003kk}
\bibinfo{author}{\bibfnamefont{T.}~\bibnamefont{Hirayama}} \bibnamefont{and}
  \bibinfo{author}{\bibfnamefont{K.}~\bibnamefont{Yoshioka}},
  \bibinfo{journal}{JHEP} \textbf{\bibinfo{volume}{01}}, \bibinfo{pages}{032}
  (\bibinfo{year}{2004}), \eprint{hep-th/0311233}.

\bibitem[{\citenamefont{Randall and
  Sundrum}(1999{\natexlab{a}})}]{Randall:1999vf}
\bibinfo{author}{\bibfnamefont{L.}~\bibnamefont{Randall}} \bibnamefont{and}
  \bibinfo{author}{\bibfnamefont{R.}~\bibnamefont{Sundrum}},
  \bibinfo{journal}{Phys. Rev. Lett.} \textbf{\bibinfo{volume}{83}},
  \bibinfo{pages}{4690} (\bibinfo{year}{1999}{\natexlab{a}}),
  \eprint{hep-th/9906064};
\bibinfo{author}{\bibfnamefont{L.}~\bibnamefont{Randall}} \bibnamefont{and}
  \bibinfo{author}{\bibfnamefont{R.}~\bibnamefont{Sundrum}},
  \bibinfo{journal}{Phys. Rev. Lett.} \textbf{\bibinfo{volume}{83}},
  \bibinfo{pages}{3370} (\bibinfo{year}{1999}{\natexlab{b}}),
  \eprint{hep-ph/9905221}.



\bibitem{tHooft:1980xb}
  G.~'t Hooft in, G.~'t Hooft  et~al (eds.)
New York, Usa: Plenum ( 1980) 438 P. ( Nato Advanced Study Institutes Series: Series B, Physics, 59)


\bibitem[{\citenamefont{Contino et~al.}(2003)\citenamefont{Contino, Nomura, and
  Pomarol}}]{Contino:2003ve}
\bibinfo{author}{\bibfnamefont{See e.g., R.}~\bibnamefont{Contino}},
  \bibinfo{author}{\bibfnamefont{Y.}~\bibnamefont{Nomura}}, \bibnamefont{and}
  \bibinfo{author}{\bibfnamefont{A.}~\bibnamefont{Pomarol}},
  \bibinfo{journal}{Nucl. Phys.} \textbf{\bibinfo{volume}{B671}},
  \bibinfo{pages}{148} (\bibinfo{year}{2003}), \eprint{hep-ph/0306259};
\bibinfo{author}{\bibfnamefont{K.}~\bibnamefont{Agashe}},
  \bibinfo{author}{\bibfnamefont{R.}~\bibnamefont{Contino}}, \bibnamefont{and}
  \bibinfo{author}{\bibfnamefont{A.}~\bibnamefont{Pomarol}},
  \bibinfo{journal}{Nucl. Phys.} \textbf{\bibinfo{volume}{B719}},
  \bibinfo{pages}{165} (\bibinfo{year}{2005}), \eprint{hep-ph/0412089};
\bibinfo{author}{\bibfnamefont{K.}~\bibnamefont{Agashe}},
  \bibinfo{author}{\bibfnamefont{R.}~\bibnamefont{Contino}},
  \bibinfo{author}{\bibfnamefont{L.}~\bibnamefont{Da~Rold}}, \bibnamefont{and}
  \bibinfo{author}{\bibfnamefont{A.}~\bibnamefont{Pomarol}},
  \bibinfo{journal}{Phys. Lett.} \textbf{\bibinfo{volume}{B641}},
  \bibinfo{pages}{62} (\bibinfo{year}{2006}), \eprint{hep-ph/0605341}.

\bibitem[{\citenamefont{Contino et~al.}(2006)\citenamefont{Contino, Da~Rold,
  and Pomarol}}]{Contino:2006qr}
\bibinfo{author}{\bibfnamefont{R.}~\bibnamefont{Contino}},
  \bibinfo{author}{\bibfnamefont{L.}~\bibnamefont{Da~Rold}}, \bibnamefont{and}
  \bibinfo{author}{\bibfnamefont{A.}~\bibnamefont{Pomarol}}
  (\bibinfo{year}{2006}), \eprint{hep-ph/0612048}.

\bibitem[{\citenamefont{Csaki et~al.}(2004{\natexlab{a}})\citenamefont{Csaki,
  Grojean, Pilo, and Terning}}]{Csaki:2003zu}
\bibinfo{author}{\bibfnamefont{See e.g., C.}~\bibnamefont{Csaki}},
  \bibinfo{author}{\bibfnamefont{C.}~\bibnamefont{Grojean}},
  \bibinfo{author}{\bibfnamefont{L.}~\bibnamefont{Pilo}}, \bibnamefont{and}
  \bibinfo{author}{\bibfnamefont{J.}~\bibnamefont{Terning}},
  \bibinfo{journal}{Phys. Rev. Lett.} \textbf{\bibinfo{volume}{92}},
  \bibinfo{pages}{101802} (\bibinfo{year}{2004}{\natexlab{a}}),
  \eprint{hep-ph/0308038};
\bibinfo{author}{\bibfnamefont{Y.}~\bibnamefont{Nomura}},
  \bibinfo{journal}{JHEP} \textbf{\bibinfo{volume}{11}}, \bibinfo{pages}{050}
  (\bibinfo{year}{2003}), \eprint{hep-ph/0309189}; 
  \bibinfo{author}{\bibfnamefont{G.}~\bibnamefont{Cacciapaglia}},
  \bibinfo{author}{\bibfnamefont{C.}~\bibnamefont{Csaki}},
  \bibinfo{author}{\bibfnamefont{G.}~\bibnamefont{Marandella}},
  \bibnamefont{and} \bibinfo{author}{\bibfnamefont{J.}~\bibnamefont{Terning}},
  \bibinfo{journal}{Phys. Rev.} \textbf{\bibinfo{volume}{D75}},
  \bibinfo{pages}{015003} (\bibinfo{year}{2007}{\natexlab{a}}),
  \eprint{hep-ph/0607146};
\bibinfo{author}{\bibfnamefont{G.}~\bibnamefont{Cacciapaglia}},
  \bibinfo{author}{\bibfnamefont{C.}~\bibnamefont{Csaki}},
  \bibinfo{author}{\bibfnamefont{G.}~\bibnamefont{Marandella}},
  \bibnamefont{and} \bibinfo{author}{\bibfnamefont{J.}~\bibnamefont{Terning}},
  \bibinfo{journal}{JHEP} \textbf{\bibinfo{volume}{02}}, \bibinfo{pages}{036}
  (\bibinfo{year}{2007}{\natexlab{b}}), \eprint{hep-ph/0611358}.
  
  \bibitem[{\citenamefont{Gripaios et~al.}(2007)}]{gmw}
\bibinfo{author}{\bibfnamefont{B.}~\bibnamefont{Gripaios}},
   \bibinfo{author}{\bibfnamefont{J.}~\bibnamefont{March-Russell}}, \bibnamefont{and}
  \bibinfo{author}{\bibfnamefont{S. ~M.}~\bibnamefont{West}},
  \eprint{Work in Progress}.
  
  \bibitem[{\citenamefont{Panico and Wulzer}(2007)}]{Panico:2007qd}
\bibinfo{author}{\bibfnamefont{G.}~\bibnamefont{Panico}} \bibnamefont{and}
  \bibinfo{author}{\bibfnamefont{A.}~\bibnamefont{Wulzer}}
  (\bibinfo{year}{2007}), \eprint{hep-th/0703287}.

  
\bibitem[{\citenamefont{Csaki et~al.}(2005)\citenamefont{Csaki, Hubisz, and
  Meade}}]{Csaki:2005vy}
\bibinfo{author}{\bibfnamefont{See e.g.,  C.}~\bibnamefont{Csaki}},
  \bibinfo{author}{\bibfnamefont{J.}~\bibnamefont{Hubisz}}, \bibnamefont{and}
  \bibinfo{author}{\bibfnamefont{P.}~\bibnamefont{Meade}}
  (\bibinfo{year}{2005}), \eprint{hep-ph/0510275}.

  

  
\bibitem[{\citenamefont{Gripaios}(2007)}]{Gripaios:2006dc}
\bibinfo{author}{\bibfnamefont{B.}~\bibnamefont{Gripaios}},
  \bibinfo{journal}{Nucl. Phys.} \textbf{\bibinfo{volume}{B768}},
  \bibinfo{pages}{157} (\bibinfo{year}{2007}), \eprint{hep-ph/0611218}.

  
  
\bibitem[{\citenamefont{Csaki et~al.}(2004{\natexlab{b}})\citenamefont{Csaki,
  Grojean, Hubisz, Shirman, and Terning}}]{Csaki:2003sh}
\bibinfo{author}{\bibfnamefont{C.}~\bibnamefont{Csaki}},
  \bibinfo{author}{\bibfnamefont{C.}~\bibnamefont{Grojean}},
  \bibinfo{author}{\bibfnamefont{J.}~\bibnamefont{Hubisz}},
  \bibinfo{author}{\bibfnamefont{Y.}~\bibnamefont{Shirman}}, \bibnamefont{and}
  \bibinfo{author}{\bibfnamefont{J.}~\bibnamefont{Terning}},
  \bibinfo{journal}{Phys. Rev.} \textbf{\bibinfo{volume}{D70}},
  \bibinfo{pages}{015012} (\bibinfo{year}{2004}{\natexlab{b}}),
  \eprint{hep-ph/0310355}.
  
  \bibitem[{\citenamefont{Hill}(2006)\citenamefont{Hill}}]{Hill:2006ei}
\bibinfo{author}{\bibfnamefont{C.~T.} \bibnamefont{Hill}},
  \bibinfo{journal}{Phys. Rev.} \textbf{\bibinfo{volume}{D73}},
  \bibinfo{pages}{085001} (\bibinfo{year}{2006}), \eprint{hep-th/0601154};
\bibinfo{author}{\bibfnamefont{C.~T.} \bibnamefont{Hill}},
  \bibinfo{journal}{Phys. Rev.} \textbf{\bibinfo{volume}{D73}},
  \bibinfo{pages}{126009} (\bibinfo{year}{2006}), \eprint{hep-th/0603060}.
  


  
\bibitem[{\citenamefont{Green and Schwarz}(1984)}]{Green:1984sg}
\bibinfo{author}{\bibfnamefont{M.~B.} \bibnamefont{Green}} \bibnamefont{and}
  \bibinfo{author}{\bibfnamefont{J.~H.} \bibnamefont{Schwarz}},
  \bibinfo{journal}{Phys. Lett.} \textbf{\bibinfo{volume}{B149}},
  \bibinfo{pages}{117} (\bibinfo{year}{1984}).
  
  
\bibitem[{\citenamefont{Groot Nibbelink}(2003)}]{GrootNibbelink:2003gb}
\bibinfo{author}{\bibfnamefont{S.} \bibnamefont{Groot Nibbelink}}, 
  \bibinfo{author}{\bibfnamefont{H.~P.} \bibnamefont{Nilles}},
  \bibinfo{author}{\bibfnamefont{M.} \bibnamefont{Olechowski}}, \bibnamefont{and}
  \bibinfo{author}{\bibfnamefont{M.~G.~A.} \bibnamefont{Walter}},
  \bibinfo{journal}{Nucl. Phys.} \textbf{\bibinfo{volume}{B665}},
  \bibinfo{pages}{236} (\bibinfo{year}{2003}), \eprint{hep-th/0303101}.
  
\bibitem[{\citenamefont{Ibanez}(1999)}]{Ibanez:1998qp}
\bibinfo{author}{\bibfnamefont{L.~E.} \bibnamefont{Ibanez}},
\bibinfo{author}{\bibfnamefont{R.} \bibnamefont{Rabadan}}, \bibnamefont{and}
  \bibinfo{author}{\bibfnamefont{A.~M.} \bibnamefont{Uranga}},
  \bibinfo{journal}{Nucl. Phys.} \textbf{\bibinfo{volume}{B542}},
  \bibinfo{pages}{112} (\bibinfo{year}{1999}), \eprint{hep-th/9808139};
  \bibinfo{author}{\bibfnamefont{C.~A.} \bibnamefont{Scrucca}} \bibnamefont{and}
  \bibinfo{author}{\bibfnamefont{M.} \bibnamefont{Serone}},
  \bibinfo{journal}{JHEP} \textbf{\bibinfo{volume}{12}},
  \bibinfo{pages}{024} (\bibinfo{year}{1999}), \eprint{hep-th/9912108};
  \bibinfo{author}{\bibfnamefont{C.~A.} \bibnamefont{Scrucca}},
 \bibinfo{author}{\bibfnamefont{M.} \bibnamefont{Serone}}, \bibnamefont{and}
  \bibinfo{author}{\bibfnamefont{M.} \bibnamefont{Trapletti}},
  \bibinfo{journal}{Nucl. Phys.} \textbf{\bibinfo{volume}{B635}},
  \bibinfo{pages}{33} (\bibinfo{year}{2002}),  \eprint{hep-th/0203190}.

\bibitem[{\citenamefont{Preskill}(1991)}]{Preskill:1990fr}
\bibinfo{author}{\bibfnamefont{J.}~\bibnamefont{Preskill}},
  \bibinfo{journal}{Ann. Phys.} \textbf{\bibinfo{volume}{210}},
  \bibinfo{pages}{323} (\bibinfo{year}{1991}).



\bibitem[{\citenamefont{Witten}(1983)}]{Witten:1983tw}
\bibinfo{author}{\bibfnamefont{J.}~\bibnamefont{Wess}} \bibnamefont{and}
  \bibinfo{author}{\bibfnamefont{B.}~\bibnamefont{Zumino}},
  \bibinfo{journal}{Phys. Lett.} \textbf{\bibinfo{volume}{B37}},
  \bibinfo{pages}{95} (\bibinfo{year}{1971});
\bibinfo{author}{\bibfnamefont{E.}~\bibnamefont{Witten}},
  \bibinfo{journal}{Nucl. Phys.} \textbf{\bibinfo{volume}{B223}},
  \bibinfo{pages}{422} (\bibinfo{year}{1983}).


\bibitem[{\citenamefont{Freedman et~al.}(1999)\citenamefont{Freedman, Mathur,
  Matusis, and Rastelli}}]{Freedman:1998tz}
\bibinfo{author}{\bibfnamefont{D.~Z.} \bibnamefont{Freedman}},
  \bibinfo{author}{\bibfnamefont{S.~D.} \bibnamefont{Mathur}},
  \bibinfo{author}{\bibfnamefont{A.}~\bibnamefont{Matusis}}, \bibnamefont{and}
  \bibinfo{author}{\bibfnamefont{L.}~\bibnamefont{Rastelli}},
  \bibinfo{journal}{Nucl. Phys.} \textbf{\bibinfo{volume}{B546}},
  \bibinfo{pages}{96} (\bibinfo{year}{1999}), \eprint{hep-th/9804058};
\bibinfo{author}{\bibfnamefont{G.}~\bibnamefont{Chalmers}},
  \bibinfo{author}{\bibfnamefont{H.}~\bibnamefont{Nastase}},
  \bibinfo{author}{\bibfnamefont{K.}~\bibnamefont{Schalm}}, \bibnamefont{and}
  \bibinfo{author}{\bibfnamefont{R.}~\bibnamefont{Siebelink}},
  \bibinfo{journal}{Nucl. Phys.} \textbf{\bibinfo{volume}{B540}},
  \bibinfo{pages}{247} (\bibinfo{year}{1999}), \eprint{hep-th/9805105}.


\bibitem[{\citenamefont{Verlinde}(2000)}]{Verlinde:1999fy}
\bibinfo{author}{\bibfnamefont{H.~L.} \bibnamefont{Verlinde}},
  \bibinfo{journal}{Nucl. Phys.} \textbf{\bibinfo{volume}{B580}},
  \bibinfo{pages}{264} (\bibinfo{year}{2000}), \eprint{hep-th/9906182};
\bibinfo{author}{\bibfnamefont{J.}~\bibnamefont{de~Boer}},
  \bibinfo{author}{\bibfnamefont{E.~P.} \bibnamefont{Verlinde}},
  \bibnamefont{and} \bibinfo{author}{\bibfnamefont{H.~L.}
  \bibnamefont{Verlinde}}, \bibinfo{journal}{JHEP}
  \textbf{\bibinfo{volume}{08}}, \bibinfo{pages}{003} (\bibinfo{year}{2000}),
  \eprint{hep-th/9912012}; 
\bibinfo{author}{\bibfnamefont{C.}~\bibnamefont{Csaki}},
  \bibinfo{author}{\bibfnamefont{J.}~\bibnamefont{Erlich}},
  \bibinfo{author}{\bibfnamefont{T.~J.} \bibnamefont{Hollowood}},
  \bibnamefont{and} \bibinfo{author}{\bibfnamefont{J.}~\bibnamefont{Terning}},
  \bibinfo{journal}{Phys. Rev.} \textbf{\bibinfo{volume}{D63}},
  \bibinfo{pages}{065019} (\bibinfo{year}{2001}), \eprint{hep-th/0003076};
\bibinfo{author}{\bibfnamefont{A.}~\bibnamefont{Lewandowski}} \bibnamefont{and}
  \bibinfo{author}{\bibfnamefont{M.}~\bibnamefont{Redi}},
  \bibinfo{journal}{Phys. Rev.} \textbf{\bibinfo{volume}{D68}},
  \bibinfo{pages}{044012} (\bibinfo{year}{2003}), \eprint{hep-th/0305013};


\bibitem[{\citenamefont{Lewandowski et~al.}(2003)\citenamefont{Lewandowski,
  May, and Sundrum}}]{Lewandowski:2002rf}
\bibinfo{author}{\bibfnamefont{A.}~\bibnamefont{Lewandowski}},
  \bibinfo{author}{\bibfnamefont{M.~J.} \bibnamefont{May}}, \bibnamefont{and}
  \bibinfo{author}{\bibfnamefont{R.}~\bibnamefont{Sundrum}},
  \bibinfo{journal}{Phys. Rev.} \textbf{\bibinfo{volume}{D67}},
 \bibinfo{pages}{024036} (\bibinfo{year}{2003}), \eprint{hep-th/0209050}.



  

\end{thebibliography}

\end{document}